\RequirePackage[2020-02-02]{latexrelease}

\documentclass[superscriptaddress,secnumarabic,
amssymb,amsmath,nobibnotes,aps,prd,showkeys,showpacs,nofootinbib]{revtex4}%
\usepackage{graphicx}
\usepackage{subfigure}
\usepackage{subcaption}
\usepackage{multirow}
\usepackage{tabularx}
\usepackage{hhline}
\usepackage{amsfonts, graphics}
\usepackage{amsmath}
\usepackage{color}
\usepackage{makecell}
\providecommand{\U}[1]{\protect\rule{.1in}{.1in}}
\newcommand{\be}{\begin{equation}}
\newcommand{\ee}{\end{equation}}

\newcommand{\mincir}{\raise
-3.truept\hbox{\rlap{\hbox{$\sim$}}\raise4.truept\hbox{$<$}\ }}
\newcommand{\magcir}{\raise
-3.truept\hbox{\rlap{\hbox{$\sim$}}\raise4.truept\hbox{$>$}\ }}

\begin{document}

\title{Designing Wormholes in Novel Power-Law $f(R)$: A Mathematical approach with a linear equation of state}
\author{Subhasis Nalui}
\email{subhasis.rs@presiuniv.ac.in }

\author{Subhra Bhattacharya}
\email{subhra.maths@presiuniv.ac.in}
\affiliation{Department of Mathematics, Presidency University, Kolkata-700073, India}


\begin{abstract}
We consider the inhomogeneous Morris-Thorne wormhole metric with matter tensors characterised by a novel linear equation of state in $f(R)$ gravity. Using the Einstein's field equations in metric $f(R)$ gravity we model solutions for both wormhole as well as $f(R)$ gravity. We obtain four different wormhole models, two wormholes are characterised by solid angle deficit, three are not asymptotically extendible, while one is asymptotically flat with zero tidal force. These are supported by four different power law $f(R)$ models. The parameter space of the models can support both null energy conditions (NEC) satisfying as well as violating wormhole. In case of NEC satisfying matter, the associated $f(R)$ is ghost. The $f(R)$ models obtained have been independently substantiated for cosmological feasibility and valid parameter space was obtained corresponding to cosmologically viable $f(R)$. Suitable scalar-tensor representation of the corresponding $f(R)$ models have been presented using the correspondence of $f(R)$ gravity with Brans-Dicke (BD) theory of gravity. The robustness of the wormhole solutions were further analysed with the BD scalar fields in the hybrid metric-Palatini gravity, which showed excellent results. Lastly as an independent astrophysical probe for the wormhole we have obtained the location of their photon spheres and have connected them with the Herrera Complexity factor in $f(R).$ Our results show that the relation between the complexity factor and existence of photon spheres remains fundamentally unaltered in $f(R)$ as compared to Einstein's gravity.    

\end{abstract}
%%%%%%%%%%%%%%%%%%%%%%%%%%%%%%%%%%%%%%%%%%%%%%%%%%%%%%%%%%%%%%%%%%%%%%%%%%%%%%%%%%%%%%%%%%%%%%%%%%%%%%%%%%%%%%
\maketitle
%%%%%%%%%%%%%%%%%%%%%%%%%%%%%%%%%%%%%%%%%%%%%%%%%%%%%%%%%%%%%%%%%%%%%%%%%%%%%%%%%%%%%%%%%%%%%%%%%%%%%%%%%%%%%%%%%%
%~~~\myclassification{98.80.Cq, 98.80.-k}\\\\
%%%%%%%%%%%%%%%%%%%%%%%%%%%%%%%%%%%%%%%%%%%%%%%%%%%%%%%%%%%%%%%%%%%%%%%%%%%%%%%%%%%%%%%%%%%%%%%%%%%%%%%%%%%%%%%%%%

\section{Introduction}

Einstein proposed General Relativity (GR) in 1915. Since then it has served as the bedrock of our understanding of the cosmos and gravity and in general providing the basis for experimentally verifiable physical theories of the universe. Despite the mathematical elegance and success of GR, it has to cope with problems of singularity, stability, absence of a committed theory of quantum gravity and finally the inexplicable dominance of recent day dark energy and dark matter driving accelerated cosmic expansion. Thus the quest for alternative theories of gravitation had been an ongoing process. The fourth order gravity theories arising out of the non-linear corrections of the curvature term in the Lagrangean, commonly known as the $f(R)$ gravity theory is one of the oldest and widely accepted theories of gravitational modification. As a mathematical curiosity, it was first explored by Herman Weyl in 1919 \cite{hw}. Interests were fuelled further when research showed that first order loop quantum corrections achieved via non-linearised curvature terms could bring about renormalization of gravity. (See \cite{sch} for a review of early historical developments of $f(R)$). This was soon followed by the Starobinsky model of inflation driven via quadratic curvature correction terms \cite{str}. Finally $f(R)$ was used as an alternative to dark energy description of the accelerated expansion of the universe \cite{defr}. This completed the cosmic description of the evolving universe using $f(R)$ gravity. However these models were riddled with problems like stability issues, inconsistency with local gravity tests and existence of ghost degrees of freedoms. Newer models and better analysis eventually eased out the complications associated with the $f(R)$ models. It was predicted to resolve the long standing coincidence problem and could prescribe a smooth evolution of the universe from the early inflation to late time acceleration with a corresponding phantom switch over, without the requirement of exotic dark components of the universe \cite{unfr}. Recent research propose methods for testing $f(R)$ using gravitational wave data \cite{dej}. (For a comprehensive review on various aspects of $f(R)$ follow \cite{far, soti, noj1, noj2, fel} and the references therein). 

Given that $f(R)$ can be an alternative to exotic matter components like dark energy and dark matter, it is natural to conceive mathematically exotic objects like wormholes within the realm of $f(R)$ gravity. Wormholes share a historical trajectory aligned with the developments of $f(R)$ gravity. Wormholes were first suggested by Flamm in 1916 \cite{flm}. Soon wormhole like solutions or intra-universe short-cut or bridges were proposed by Einstein and Rosen in 1935 \cite{er,vis}. Finally the closure of twentieth century saw a renewed impetus in wormhole research on account of the Morris-Thorne traversable wormhole \cite{mt}. In \cite{mt, mty} a traversable wormhole was contrived to be a time-machine useful for inter-universe/inter-galactic travels. However such engineering compromised with the energy conditions of the resulting matter tensors making them exotic \cite{mt,hv}. Research on wormholes were mostly targeted to mollify the violation of the energy conditions of the matter tensors.  Modified gravity was proposed as a remedy for null energy condition violating matter tensors. 
Here we shall construct the Morris-Thorne traversable wormhole in the background of the metric $f(R)$ gravity by assuming a phenomenological linear equation of state (eos) given by 
\begin{equation}
\rho=-\frac{p_{r}+\delta p_{t}}{1+\delta},~ (\delta\neq -1).\label{eos}
\end{equation}
The choice of the above eos will be useful to construct the master equation from the Einstein field equations for evaluating the modelled system. Additionally the eos of the matter tensors impact their energy conditions hence it is natural that, in traversable wormhole mathematics they will play some significant part. In this context, we note that the eos (\ref{eos}) could be written as $\rho+p_{r}=-\delta(\rho+p_{t}),$ that is, one can obtain the NEC directly from the above eos with NEC always violated for a positive $\delta,$ while NEC can be satisfied for a negative $\delta$ and for $\delta=0$ one can obtain the cosmological constant. In past similar equation of states have been used in the context of higher gravity theories and wormholes. In \cite{abt} the eos $\gamma \rho=p_{r}+2p_{t}$ was used to obtain NEC satisfying wormhole in the Brans-Dicke gravity. This eos of state was modified to $\rho=\omega[p_{r}+(n-2)p_{t}]$ for the $n$ dimensional wormhole geometry in second and third order Lovelock gravity \cite{reza1,reza2}. More recently this has been used for wormholes in $f(Q)$ gravity \cite{kiro}, while in \cite{zub} the eos $\gamma \rho=p_{r}+2p_{t}$ has been applied for wormholes in four dimensional Gauss-Bonnet gravity. 

Our aim is to develop suitable models for wormhole and $f(R)$ gravity guided by the above eos. Here we shall not assume any existing $f(R)$ or wormhole model, instead we shall design the models via mathematical solution that are generic to the above eos. Since the above eos can also represent the NEC satisfying system, we shall explore all such possibilities where such systems can be obtained. We shall also explore the constructed $f(R)$ models for their cosmological viability and the wormhole models for their astrophysical significance by determining the location of their photon spheres. 

The article is arranged as follows: In section 2 we shall describe the background metric with the $f(R)$ action. In section 3 we shall formulate the master equation and provide corresponding solution method for solving the master equation. In subsection 1 and 2 of section 3, we will furnish the mathematical details and physical relevance of the evaluated solutions. In section 4 we give the details of the energy conditions satisfied by the matter lining the wormhole throat and its corresponding consequence of $f(R)$ gravity. Using the notion of interchangeability of $f(R)$ gravity and scalar-tensor BD gravity we provide the scalar field and scalar potential perspective of the $f(R)$ gravity solutions in section 5. In subsection 1 we shall carry out an independent analysis of the $f(R)$ models for their cosmological viability. In section 6 we will find the locations of the photon spheres of the wormhole solutions and show how they can be obtained easily using the complexity of the wormhole for any radius in the vicinity of the wormhole throat. Finally in section 7 we present the conclusion.

\section{Morris-Thorne wormhole in the background of Metric $f(R)$ gravity}

$f(R)$ gravity has three versions, the metric $f(R)$, the Palatini $f(R)$ and Metric-affine $f(R)$ gravity. Here we shall model the metric $f(R),$ which was studied by Buchdhal \cite{buch} in 1970. In metric formalism the action for $f(R)$ is given by:
\begin{equation}
S_{f}=\frac{1}{2\kappa}\int f(R)\sqrt{-g}d^{4}x.\label{faction}
\end{equation}
($\kappa=\frac{1}{8\pi G}$ is the Einstein's gravitational constant). The effective gravitational action is given as:
\begin{equation}
S=S_{f}+S_{M}\label{taction}
\end{equation}
where $S_{M}$ is the action due to matter fields. Varying the action with respect to (wrt) the metric $g_{\mu\nu}$ we get the field equations as:
\begin{equation}
f_{R}R_{\mu\nu}+\left(g_{\mu\nu}\square-\triangledown_{\mu}\triangledown_{\nu}\right)f_{R}-\frac{1}{2}g_{\mu\nu}f=\kappa T_{\mu\nu}\label{ffe}
\end{equation}
where $f_{R}=\frac{d f(R)}{d R},~\square$ is the d'Alembertian operator, $\triangledown_{\mu}$ the covariant derivative, and $T_{\mu\nu}$ is the matter tensors arising from the variation of $S_{M}$ which can be described using the equation
\begin{equation}
T_{\mu\nu}=(\rho+p_{t})u_{\mu}u_{\nu}+p_{t}g_{\mu\nu}+(p_{r}-p_{t})X_{\mu}X_{\nu}
\end{equation}
where $\rho$ is the energy density, $p_{r}$ and $p_{t}$ are the radial and tangential pressure respectively. $u^{\mu}$ is the four velocity while $X^{\mu}$ is the unit space like vector orthogonal to the four velocity. We will consider the field equation (\ref{ffe}) in a static traversable wormhole geometry expressed by the metric:
\begin{equation}
dS^{2}=-e^{2\phi(r)}dt^{2}+\frac{dr^{2}}{1-b(r)/r}+r^{2}(d\theta^{2}+\sin^{2}\theta d\psi^{2})\label{metric}
\end{equation} 
where the function $\phi(r)$ and $b(r)$ are the redshift function and the shape function respectively. A traversable wormhole does not have a horizon, that is $\phi(r)$ is finite for all values of $r.$ Here $r$ can range from a minimum $r_{0}$ to $\infty.$ The radius $r_{0}$ is the radius of a sphere located at the wormhole throat. The shape of the throat is characterized by the function $b(r)$ such that $1-\frac{b(r)}{r}\geq 0$ for $r\geq r_{0}$ with equality obtained only at $r_{0}.$ The function $b(r)$ should also satisfy the flare out condition at the throat given by $b'(r_{0})<1.$ Further the above geometry will be asymptotically flat provided $\frac{b(r)}{r}\rightarrow 0$ as $r\rightarrow\infty.$ As already noted, a direct consequence of the properties of the throat is that the throat is supported by NEC violating matter or ``exotic" matter, where radial stress exceeds the matter density. $f(R)$ gravity is looked upon as a useful mechanism for avoiding the exotic matter. Several literatures show that the $f(R)$ induced modification in the field equations can result in the existence of the wormhole with matter tensors satisfying NEC \cite{fr}. However this apparent nicety comes with a caveat, that the associated graviton field is rendered ghost \cite{bron}. 

The Ricci curvature $R$ in the above metric is given as:
\begin{equation}
R(r)=-2\left(1-\frac{b(r)}{r}\right)\left(\phi''(r)+(\phi'(r))^{2}+\frac{2\phi'(r)}{r}\right)+\left(\frac{rb'(r)-b}{r^{2}}\right)\phi'(r)+\frac{2b'(r)}{r^{2}},\label{R}
\end{equation}
while the field equations in $f(R)$ and with wormhole geometry is explicitly given by:
\begin{align}
\kappa \rho(r)=&\frac{1}{2}f(R)+\left(\frac{b'(r)}{r^{2}}-\frac{R}{2}\right)F(R)+\left[\left(\frac{rb'(r)-b}{2r^{2}}\right)-\frac{2}{r}\left(1-\frac{b(r)}{r}\right)\right]F'(R)-\left(1-\frac{b(r)}{r}\right)F''(R)\label{fe1}\\
 \kappa p_{r}(r)=&-\frac{1}{2}f(R)+\left[\frac{2}{r}\left(1-\frac{b(r)}{r}\right)\phi'(r)-\frac{b(r)}{r^{3}}+\frac{R}{2}\right]F(R)+\left(1-\frac{b(r)}{r}\right)\left(\frac{2}{r}+\phi'(r)\right)F'(R)\label{fe2}\\
\kappa p_{t}(r)=&-\frac{1}{2}f(R)+\left[\left(1-\frac{b(r)}{r}\right)\left(\phi''(r)+(\phi'(r))^{2}+\frac{\phi'(r)}{r}\right)-\left(\frac{rb'(r)-b}{2r^{2}}\right)\left(\phi'(r)+\frac{1}{r}\right)+\frac{R}{2}\right]F(R)\nonumber  \\ 
&+\left[\left(\phi'(r)+\frac{1}{r}\right)\left(1-\frac{b(r)}{r}\right)-\left(\frac{rb'(r)-b}{2r^{2}}\right)\right]F'(R)+\left(1-\frac{b(r)}{r}\right)F''(R)\label{fe3}
\end{align}
where {\it prime} denotes derivative wrt $r$ and $F(R)=f_{R}(R).$ Since the curvature is a function of $r$, we obtain the $f(R)$ as a function of $r$ that is we will obtain $f(R(r))$ and consequently $F(R(r)).$ 

\section{$f(R)$ supported wormhole geometry}

Using the equation of state (\ref{eos}) we get a relation between the functions $f(R), ~b(r)$ and $\phi(r).$ The corresponding equation is a differential equation in three unknown functions $f(R), ~b(r)$ and $\phi(r).$ Since relation (\ref{eos}) connects the three equations (\ref{fe1})-(\ref{fe3}), the resulting equation will have linear first order differential terms for the shape function $b(r)$, quadratic first order and linear second order differential terms of $\phi(r)$ and linear second order differential terms of $F(R).$ In the existing literature such equations are usually solved by assuming known forms of any two functions and then solving for the third. Here we shall employ a method such that we can solve for all three functions in order, instead of assuming a specific form for any of them. 

In order to simplify the process we write the equation as a first order equation in $\frac{b(r)}{r}$ as follows: 
\begin{equation}
\frac{d}{dr}\left(\frac{b(r)}{r}\right)+\xi(r)\left(\frac{b(r)}{r}\right)=\eta(r)\label{me}
\end{equation}
where 
\begin{align}
\xi(r)=&\frac{F''(R)-\left((1+\delta)\phi'(r)-\frac{\delta}{r}\right)F'(R)+\left[\frac{\delta}{r^{2}}-\delta(\phi''(r)+(\phi'(r))^{2})-\left(\frac{2+\delta}{r}\right)\phi'(r)\right]F(R)}{\frac{F'(R)}{2}+\left[\left(\frac{2+\delta}{2r}\right)-\left(\frac{\delta\phi'(r)}{2}\right)\right]F(R)}\\
\eta(r)=&\frac{F''(R)-\left((1+\delta)\phi'(r)-\frac{\delta}{r}\right)F'(R)-\left[\delta(\phi''(r)+(\phi'(r))^{2})+\left(\frac{2+\delta}{r}\right)\phi'(r)\right]F(R)}{\frac{F'(R)}{2}+\left[\left(\frac{2+\delta}{2r}\right)-\left(\frac{\delta\phi'(r)}{2}\right)\right]F(R)}
\end{align}
We can solve (\ref{me}) for any $\xi(r)$ and $\eta(r)$ as:
\begin{equation}
\frac{b(r)}{r}=e^{-\int\xi(r)dr}\left[\int\eta(r)e^{\int\xi(r)dr}dr+b_{0}\right]\label{b}
\end{equation} 
where $b_{0}$ is the constant of integration. We will require specific functional forms of the coefficients $\xi(r)$ and $\eta(r)$ that are integrable and satisfy the wormhole throat restrictions. In order to reduce $\xi$ and $\eta$ to viable integrable forms we shall assume constraints on the functions $\phi(r)$ and $F(R).$ Accordingly we consider: 
\begin{equation}
\delta(\phi''(r)+(\phi'(r))^{2})+\left(\frac{2+\delta}{r}\right)\phi'(r)=0\label{phi}
\end{equation}
The solution to this gives us the red-shift function $e^{\phi(r)}=\phi_{1}-\left(\frac{\delta\phi_{0}}{2}\right)r^{-\frac{2}{\delta}}$ with $\phi_{0}$ and $\phi_{1}$ being constants of integration. Here for $\delta>0,~e^{\phi(r)}\rightarrow\phi_{1}$ as $r\rightarrow\infty$ while for $\delta<0,~e^{\phi(r)}\rightarrow\infty$ as $r\rightarrow\infty.$   Next we assume: 
\begin{equation}
F''(R)-\left((1+\delta)\phi'(r)-\frac{\delta}{r}\right)F'(R)=f_{1}r^{\alpha}\label{F}
\end{equation}
where $f_{1}$ is a constant parameter and $\alpha$ is any real constant. Analytic solution of this equation requires either $(\phi_{1}=0,\phi_{0}\neq 0)$ or $(\phi_{1}\neq 0, \phi_{0}=0).$ The second case reduces to the scenario of zero-tidal force wormholes. 

\subsection{Non-asymptotic wormhole in power law $f(R)$ for $\phi_{1}=0$}

First set of solutions are obtained with $\phi_{1}=0,
~(\delta,\phi_{0})\neq 0$ as:
\begin{equation}
F(r)=F_{1}(\delta,\alpha)r^{\alpha+2}+F_{2}(\delta,\alpha)r^{-\frac{\delta^{2}+\delta+2}{\delta}}+F_{3}(\delta,\alpha)\label{F1}
\end{equation}
with 
\begin{align}
F_{1}(\delta,\alpha)=&f_{1}\delta\left[(\alpha+2)(\delta^{2}+\delta(3+\alpha)+2)\right]^{-1},~\alpha\neq -2~\text{and}~\alpha\neq -\frac{\delta^{2}+3\delta+2}{\delta}\\
F_{2}(\delta,\alpha)=&f_{2}\delta\left[\delta^{2}+\delta+2\right]^{-1},~\delta^{2}+\delta+2\neq 0\\
F_{3}(\delta,\alpha)=&f_{3}
\end{align}
$f_{2}$ and $f_{3}$ being constants of integration. (It may be noted that Einstein's GR can be obtained for $f_{1}=f_{2}=0$ and $f_{3}=1$). $\xi(r)$ and $\eta(r)$ reduces to non-trivial integrable form for either $f_{1}=0$ or $f_{2}=0$ and $f_{3}=0.$ 

\begin{itemize}
\vspace{1em}
\item {\bf Solution 1:}\\

 Let $f_{2}=0,~f_{3}=0.$ This gives $\xi(r)=\frac{\xi_{1}}{r}$ and $\eta(r)=\frac{\eta_{1}}{r}$ where 
\begin{align*}
\xi_{1}=&\frac{2(\alpha+2)(\delta^{2}+(3+\alpha)\delta+2)+2\delta^{2}}{\delta(\alpha+\delta+6)}\\
\eta_{1}=&\frac{2(\alpha+2)(\delta^{2}+(3+\alpha)\delta+2)}{\delta(\alpha+\delta+6)}
\end{align*}  
With $\delta\ne 0,~\alpha\neq -(\delta+6).$ Using this $\xi(r)$ and $\eta(r)$ in equation (\ref{b}) together with the wormhole throat condition we get 
\begin{equation}
\frac{b(r)}{r}=\frac{\eta_{1}}{\xi_{1}}+\left(1-\frac{\eta_{1}}{\xi_{1}}\right)\left(\frac{r_{0}}{r}\right)^{\xi_{1}}\label{br1}
\end{equation} 
where $r_{0}$ is the location of the wormhole throat. Imposing the wormhole throat flare out condition we get that $0\leq \frac{\eta_{1}}{\xi_{1}}<1,~\xi_{1}>0.$ The Ricci scalar curvature $R$ can be obtained from equation (\ref{R}) using the above forms of $b(r)$ and $\phi(r)$ as:
\begin{equation}
R(r)=\left(\frac{r_{0}}{r}\right)^{2}\left[R_{1}+R_{2}\left(\frac{r_{0}}{r}\right)^{\xi_{1}}\right]\label{R1}
\end{equation} 
where 
\begin{align*}
R_{1}=&\frac{2}{r_{0}^{2}}\left[\frac{2(\delta-2)}{\delta^{2}}\left(1-\frac{\eta_{1}}{\xi_{1}}\right)+\left(\frac{\eta_{1}}{\xi_{1}}\right)\right]\\
R_{2}=&\frac{2}{r_{0}^{2}}\left(1-\frac{\eta_{1}}{\xi_{1}}\right)\left[\left(\frac{1-\delta}{\delta}\right)\xi_{1}+\left(\frac{\delta^{2}-2\delta+4}{\delta^{2}}\right)\right]
\end{align*}
Using this $R$ we can obtain $f(R(r))$ as follows:
\begin{equation}
f(R(r))=-F_{01}\left(\frac{r_{0}}{r}\right)^{-\alpha}\left[R_{1}\frac{2}{\alpha}+R_{2}\frac{(\xi_{1}+2)}{\alpha-\xi_{1}}\left(\frac{r_{0}}{r}\right)^{\xi_{1}}\right]\label{fr1}
\end{equation}
where $F_{01}=F(R(r_{0}))=F_{1}(\delta,\alpha)r_{0}^{\alpha+2}$ is the value of $F(r)$ at the throat. With the above solutions the wormhole metric is given by:
\begin{equation}
dS^{2}=-\left(\frac{\delta\phi_{0}}{2}\right)^{2}r^{-\frac{4}{\delta}}dt^{2}+\left(1-\frac{\eta_{1}}{\xi_{1}}\right)^{-1}\frac{dr^{2}}{1-\left(\frac{r_{0}}{r}\right)^{\xi_{1}}}+r^{2}(d\theta^{2}+\sin^{2}\theta d\psi^{2})\label{met1}
\end{equation} 
For the validity of the wormhole (\ref{met1}) we had imposed various parameter restrictions on equations (\ref{F1}) and (\ref{br1}) which summarizes as the following validity ranges on $\alpha$ and $\delta$:
\begin{description}
\item Region I: $\delta<-4$ and $-2<\alpha<-(\delta+6).$ 
\item Region II: $0<\delta\leq\frac{2}{3}$ and $-2<\alpha<\infty$ 
\item Region III: $\delta>\frac{2}{3}$ and $\alpha\in(-2,\infty)\cup(-\delta-6,-\delta-3-\frac{2}{\delta})$
\end{description}

The wormhole (\ref{met1}) has an additional factor $\left(1-\frac{\eta_{1}}{\xi_{1}}\right)^{-1}$ associated with the shape function. This can be fixed by redefining the radial coordinate as $r^{2}=\left(1-\frac{\eta_{1}}{\xi_{1}}\right)r_{new}^{2}.$ However, the geometry in the new radial coordinate will have a solid angle deficit of $\left(\frac{\eta_{1}}{\xi_{1}}\right)<1.$ Similar wormholes have been previously studied in the context of general relativity in   \cite{lob1,cat1,cat2}. These wormholes are not asymptotically flat because $\frac{b(r)}{r}\rightarrow\frac{\eta_{1}}{\xi_{1}}(\neq 0)$ as $r\rightarrow\infty.$ In the parameter validity {\it Region I}, which exits for $\delta<0$ we observe that the red-shift function diverges as $r\rightarrow\infty,$ showing that in this region all conditions for asymptotic flatness are violated. Such wormholes are finite sized and can be defined for some finite value of $r=r_{max}(>r_{0}).$ In GR similar wormhole geometry can be described corresponding to isotropic pressure \cite{lob1,cat1}. The parameter validity {\it Region II} and {\it III} exits for $\delta>0$, and for red-shift function $e^{\phi(r)}\rightarrow 0$ as $r\rightarrow\infty.$ But again this is not a problem because the wormhole in this solution is not asymptotically flat, (due to solid angle deficit) and exits only for some finite $r=r_{max}(>r_{0}).$ The geometry as $r\rightarrow\infty$ can be defined by trimming the wormhole at $r_{max}$ and matching it with an exterior Schwarzschild vacuum space-time. The Schwarzschild mass $M=\frac{b(r_{max})}{2}$ is equivalent to the wormhole mass at the junction. In GR finite sized wormholes have been previously studied in various context in the literature \cite{lob1,cat1,cat2,cat3,sg,rah2}. Our analysis provides a mathematical mechanism of obtaining finite sized non-asymptotically flat wormholes.

\vspace{1em}
\item {\bf Solution 2:}\\

Let $f_{1}=0,~f_{3}=0.$ This gives $\xi(r)=\frac{\xi_{2}}{r}$ and $\eta(r)=0,$ with $\xi_{2}= \frac{2\delta^{2}}{3\delta-2}.$ The shape function is given by:
\begin{equation}
\frac{b(r)}{r}=\left(\frac{r_{0}}{r}\right)^{\xi_{2}}\label{br2}
\end{equation}
For the existence of wormhole throat and its flaring out, we require $\xi_{2}>0.$ The Ricci scalar $R(r)$ for this is given by:
\begin{equation}
R(r)=\left(\frac{r_{0}}{r}\right)^{2}\left[R_{3}+R_{4}\left(\frac{r_{0}}{r}\right)^{\xi_{2}}\right]\label{R2}
\end{equation}
where 
\begin{align*}
R_{3}=&\frac{4(\delta-2)}{\delta^{2}r_{0}^{2}}\\
R_{4}=&\frac{2}{r_{0}^{2}}\left[\xi_{2}\left(\frac{1-\delta}{\delta}\right)+\left(\frac{\delta^{2}-2\delta+4}{\delta^{2}}\right)\right]
\end{align*}
This gives the following $f(R(r))$ description:
\begin{equation}
f(R(r))=F_{02}\left(\frac{r_{0}}{r}\right)^{\beta+2}\left[R_{3}\frac{2}{\beta+2}+R_{4}\frac{(\xi_{2}+2)}{\beta+2+\xi_{2}}\left(\frac{r_{0}}{r}\right)^{\xi_{2}}\right]\label{fr2}
\end{equation}
where $\beta=\frac{\delta^{2}+\delta+2}{\delta}$ and $F_{02}=F(R(r_{0}))=F_{2}(\delta,\alpha)r_{0}^{-\beta}.$ Since the solutions (\ref{br2}-\ref{fr2}) are independent of parameter $\alpha,$ the wormhole depends on the parameter $\delta$ and exists for $\delta>\frac{2}{3}.$ 

Unlike the shape function of Solution 1, the above described wormhole geometry is solid-angle complete with asymptotic flatness condition satisfied. However, since the wormhole exists for $\delta>0$, the red-shift function $e^{\phi(r)}\rightarrow 0$ as $r\rightarrow\infty,$ which means that this wormhole  can develop an event horizon at infinite radius, making it non-traversable. Moreover, existence of event horizon at infinite radius is not a physically viable attribute of any space-time, so this wormhole cannot be defined in the asymptotically flat sense. Instead we again define a finite wormhole upto some radius $r_{0}<r<r_{max},$ as in Solution 1, with a matching Schwarzschild vacuum solution for $r\geq r_{max}.$
\end{itemize}

The corresponding $f(R)$ in both solutions are power-law in $R.$ Although we have not expressed the $f(R)$ models (\ref{fr1}) and (\ref{fr2}) explicitly using $R,$ but that they are power law in $R$ can be reasoned from the expressions of the scalar curvature given by (\ref{R1}) and (\ref{R2}). Such power-law $f(R)$ models have been recently studied quite extensively in the literature. Cosmology using a combination of power law $f(R)$ shows various promising features and have been used to describe a unified cosmology from early inflation to late time dark energy era \cite{odi1,odi2,odi3,odi4,odi5}. Power law $f(R)$ have been used in describing an anisotropic universe \cite{sham1}, in the context of gravity waves \cite{gog} and in describing inflationary cosmology \cite{inf}. Given the variety of applications of the power-law $f(R)$ that are already in the literature, our solutions become greatly significant. Here we have suggested a physically and mathematically viable mechanism for the natural occurrence of power-law $f(R)$ model in an inhomogeneous universe. We emphasise that both wormhole and $f(R)$ models proposed in Solutions 1 and 2 are obtained as viable solutions of the eos model and not constructed by choice, as is usually done in the literature. Although, our models have some severe parameter dependence, this can be relaxed while considering the $f(R)$ models independently of their wormhole background.

In FIG. 1 we have provided a graphical representation of the wormhole embedding function for Solutions 1 and 2 along with the $f(R)$ solutions (\ref{fr1}) and (\ref{fr2}) as a function of the curvature $R$ as obtained in (\ref{R1}) and (\ref{R2}). The sub-figure 1(a) clearly shows that the wormhole shape function in Solution 1 cannot become asymptotically flat. The parameters selected corresponding to {\it Region I} of the parameter space. The Sub-figure 1(c) is correspond to the wormhole shape function (\ref{br2}) which will be asymptotically flat as $r$ increases. However it may be reminded that the wormhole in Solution 2 was defined as a finite wormhole due to the physically anomalous behaviour of the red-shift function as $r\rightarrow\infty.$

\begin{figure}
\centering
\subfigure[]{\includegraphics[width=7cm,height=5cm]{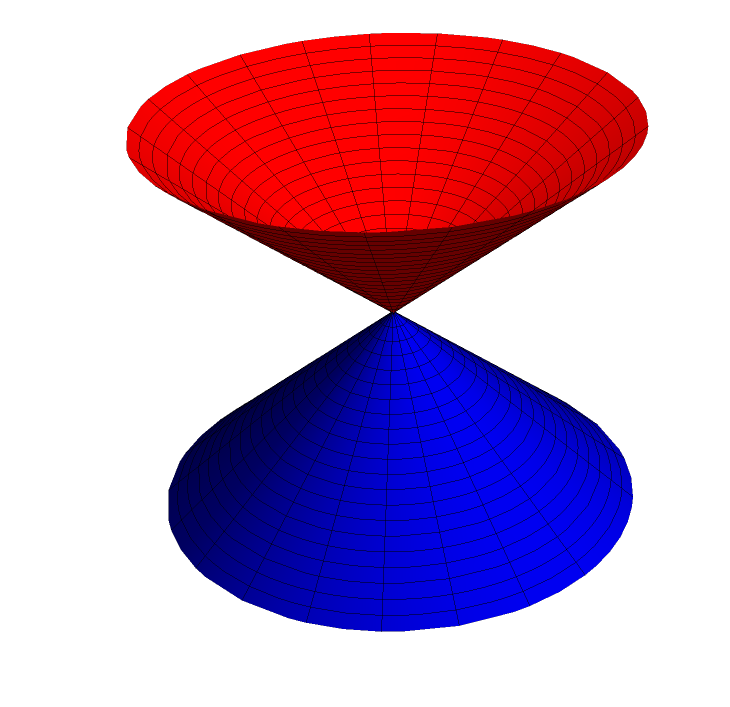}}\hspace{2cm}
\subfigure[]{\includegraphics[width=7cm,height=5cm]{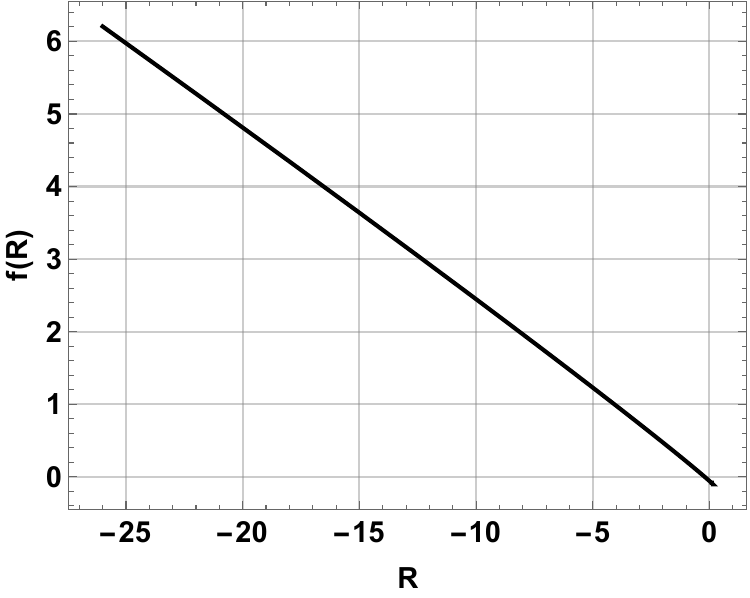}}\\
\subfigure[]{\includegraphics[width=7cm,height=5cm]{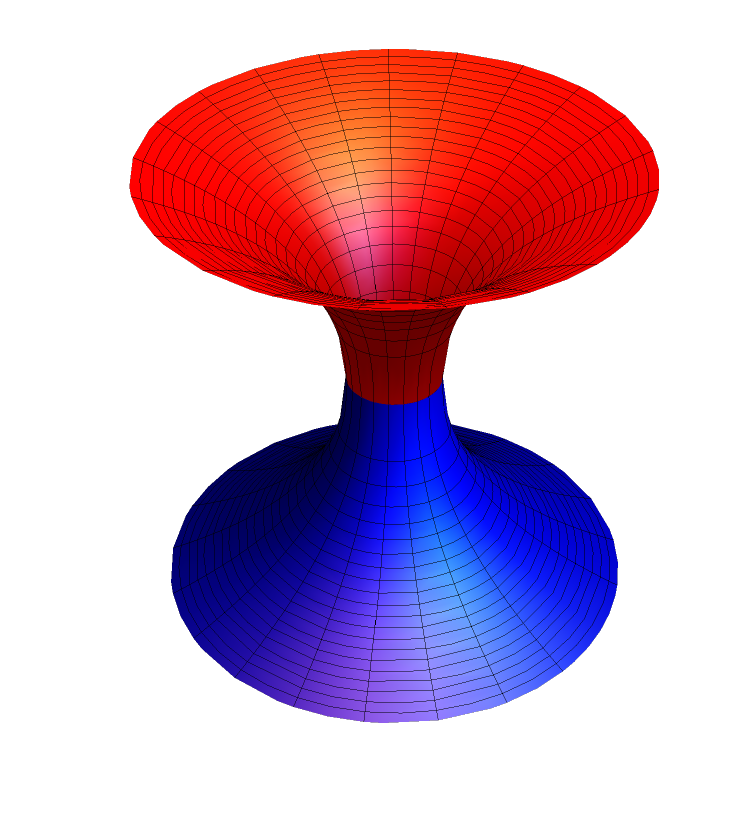}}\hspace{2cm}
\subfigure[]{\includegraphics[width=7cm,height=5cm]{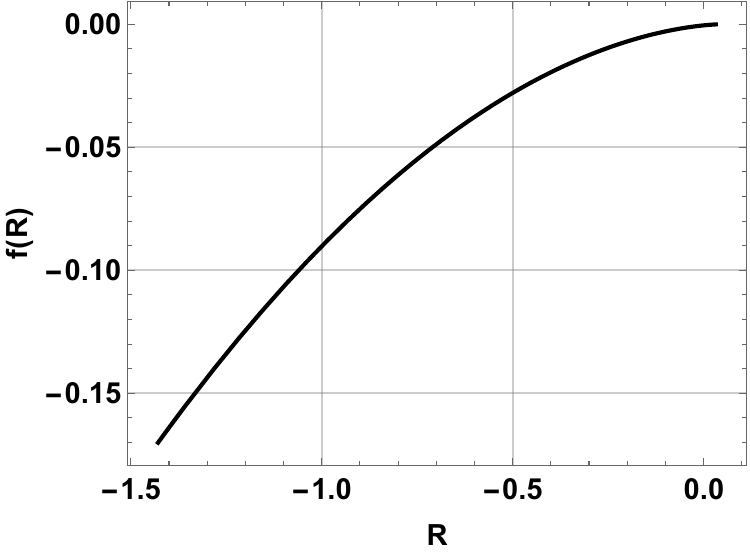}}
\caption{Figures (a) and (b) respectively depict the embedding diagram for the wormhole throat and the $f(R)$ as a function of the curvature $R$ corresponding to Solution 1. The figures have been obtained for $r_{0}=1,~\delta=-6,~\alpha=-1$ and $f_{1}=1.$ Figures (c) and (d) respectively show the embedding for the wormhole throat and the $f(R)$ solution as a function of $R$ corresponding to Solution 2. Both (c) and (d) have been constructed with parameters $r_{0}=1,~\delta=3,~f_{2}=1.$}
\label{fig:Fone}
\end{figure}

\subsection{Zero-tidal force wormhole in power-law $f(R)$ for $\phi_{0}=0$}

The second set of solutions obtained for $\phi_{0}=0$ or the class of zero-tidal force wormhole as:
\begin{equation}
F(r)=F_{4}(\delta,\alpha)r^{\alpha+2}+F_{5}(\delta,\alpha)r^{1-\delta}+F_{6}(\delta,\alpha)\label{F2}
\end{equation}
where 
\begin{align}
F_{4}(\delta,\alpha)=&f_{1}\left[(\alpha+2)(\alpha+\delta+1)\right]^{-1},~\alpha\neq -2~\text{and}~\alpha\neq -(1+\delta)\\
F_{5}(\delta,\alpha)=&f_{4}(1-\delta)^{-1},~\delta\neq 1\\
F_{6}(\delta,\alpha)=&f_{5}
\end{align}
$f_{4}$ and $f_{5}$ being constants of integration. Following the procedure as mentioned above we get:

\begin{itemize}
\item {\bf Solution 3:}\\

 Let $f_{4}=0,~f_{5}=0.$ This gives $\xi(r)=\frac{\xi_{3}}{r}$ and $\eta(r)=\frac{\eta_{3}}{r}$ where 
\begin{align*}
\xi_{3}=&\frac{2(\alpha+2)(\alpha+\delta+1)+2\delta}{\alpha+\delta+4}\\
\eta_{3}=&\frac{2(\alpha+2)(\alpha+\delta+1)}{\alpha+\delta+4}
\end{align*}  
As before we get 
\begin{equation}
\frac{b(r)}{r}=\frac{\eta_{3}}{\xi_{3}}+\left(1-\frac{\eta_{3}}{\xi_{3}}\right)\left(\frac{r_{0}}{r}\right)^{\xi_{3}}\label{br3}
\end{equation} 
where $r_{0}$ is the location of the wormhole throat. Imposing the wormhole throat flare out condition we get that $0\leq \frac{\eta_{3}}{\xi_{3}}<1,~\xi_{3}>0.$ The Ricci scalar curvature $R$ can be obtained from equation (\ref{R}) using the above form of $b(r)$ as:
\begin{equation}
R(r)=\left(\frac{r_{0}}{r}\right)^{2}\left[R_{5}+R_{6}\left(\frac{r_{0}}{r}\right)^{\xi_{3}}\right]\label{R3}
\end{equation} 
where 
\begin{align*}
R_{5}=&\frac{2}{r_{0}^{2}}\left(\frac{\eta_{3}}{\xi_{3}}\right)\\
R_{6}=&\frac{2}{r_{0}^{2}}\left(1-\frac{\eta_{3}}{\xi_{3}}\right)(1-\xi_{3})
\end{align*}
Thus we can obtain $f(R(r))$ as follows:
\begin{equation}
f(R(r))=-F_{03}\left(\frac{r_{0}}{r}\right)^{-\alpha}\left[R_{5}\frac{2}{\alpha}+R_{6}\frac{(\xi_{3}+2)}{\alpha-\xi_{3}}\left(\frac{r_{0}}{r}\right)^{\xi_{3}}\right]\label{fr3}
\end{equation}
where $F_{03}=F(R(r_{0}))=F_{4}(\delta,\alpha)r_{0}^{\alpha+2}$ is the value of $F(r)$ at the throat.
The metric representation corresponding to the above solutions is given by:
\begin{equation}
dS^{2}=-dt^{2}+\left(1-\frac{\eta_{3}}{\xi_{3}}\right)^{-1}\frac{dr^{2}}{1-\left(\frac{r_{0}}{r}\right)^{\xi_{3}}}+r^{2}(d\theta^{2}+\sin^{2}\theta d\psi^{2})\label{met3}
\end{equation}
(Here we assumed $e^{2\phi(r)}=\phi_{1}^{2}=1$). Parameter space where the corresponding wormhole solution is valid can be identified as:
\begin{description}
\item Region IV: $\delta<-2$ and $\alpha\in(-2,-(\delta+4)).$
\item Region V: $\delta\geq 1$ and $\alpha\in(-2,\infty)\cup(-(\delta+4),-(\delta+1))$
\end{description}
Again we have a wormhole for which asymptotic flatness condition of the shape function is violated due to a solid angle deficit of $\frac{\eta_{3}}{\xi_{3}}$. Here the red-shift function $e^{\phi(r)} =\phi_{1}$ is a constant, with such wormholes exercising no tidal forces. As in the previous two wormhole solutions, this is also finite sized with $r_{0}<r<r_{max}$ with the geometry exterior to $r_{max}$ described by the Schwarzschild vacuum solution.

 \vspace{1em}
\item {\bf Solution 4:}\\ 

Let $f_{1}=0,~f_{5}=0.$ This gives $\xi(r)=\frac{\xi_{4}}{r}$ and $\eta(r)=0$ and hence
\begin{equation}
\frac{b(r)}{r}=\left(\frac{r_{0}}{r}\right)^{\xi_{4}}\label{br4}
\end{equation}
where $\xi_{4}=\frac{2\delta}{3}.$ To satisfy the wormhole throat and its flaring out $\xi_{4}>0\Rightarrow\delta>0.$ The Ricci scalar $R(r)$ for this is given by:
\begin{equation}
R(r)=R_{7}\left(\frac{r_{0}}{r}\right)^{(\xi_{4}+2)}\label{R4}
\end{equation}
where $R_{7}= \frac{2}{r_{0}^{2}}(1-\xi_{4}).$ This gives the following $f(R(r))$ gravity description:
\begin{equation}
f(R(r))=F_{04}R_{7}\left(\frac{\xi_{4}+2}{\xi_{4}+\delta+1}\right)\left(\frac{r_{0}}{r}\right)^{\xi_{4}+\delta+1}\label{fr4}
\end{equation}
where $F_{04}=F(R(r_{0}))=F_{5}(\delta,\alpha)r_{0}^{1-\delta}.$
The traversable wormhole solution obtained is asymptotically flat with zero-tidal forces. These are the most common type of wormholes that can be found in the literature and have been extensively studied in various context and in various gravity theories.
\end{itemize}

The $f(R)$ in Solution 3 given by equation (\ref{fr3}) is similar to those already obtained in Solutions 1 and 2. The $f(R)$ model obtained in Solution 4, although power-law has some additional interesting features. Firstly, in this solution it is possible to have explicit analytical expression for $f(R)$ solution (\ref{fr4}) as: $f(R)=F_{04}R_{7}\left(\frac{2\delta+6}{5\delta+3}\right)\left(\frac{R}{R_{7}}\right)^{1+\epsilon}$ where $\epsilon=\left(\frac{3\delta-3}{2\delta+6}\right).$ For small values of $\epsilon$ (for the above $f(R)$ model $\epsilon<1$ for $-3<\delta<9$ and $\epsilon\ll 1$ for $\delta$ near to 1), one can expand $f(R)\simeq R+\epsilon R\log R+O(\epsilon^{2}).$ This model of $f(R)$ have been previously studied in various contexts, like in cosmology \cite{cb}, for tuning cosmological background of gravity waves during inflation \cite{clf}, for obtaining spherically and axially symmetric solutions using Noether symmetry \cite{cap1}, to test gravitational theories using eccentric eclipsing detached binaries \cite{mrf} and for explaining observed mass relation in neutron stars and binary mergers \cite{cap2}.  In \cite{cap3} the above $f(R)$ model was used to obtain stable traversable wormhole without exotic matter. The stability of the wormhole in \cite{cap3} was established by imposing the causal restrictions on the squared adiabatic sound speed $c_{s}^{2}=\frac{\partial p}{\partial \rho},$ interpreted as the speed of propagation of perturbations in an isotropic homogeneous fluid. Such isotropization of the fluid in the wormhole context is done by taking the average pressure as $\frac{p_{r}+2p_{t}}{3}$ \cite{cap3,cap5}. In our study, by the choice of the equation of state (\ref{eos}), isotropization is inherent if one chooses the average pressure $p_{avg}(\delta)=-\frac{p_{r}+\delta p_{t}}{1+\delta}.$ As a result for any $\delta$ (including $\delta=2$) we get $c_{s}^{2}= 1$ and $\rho=p_{avg}.$ This means that the wormhole is always supported by isotropized stiff matter, which is luminal without violating the causality conditions. If we analyse the isotropized NEC $\rho+p_{avg}(\delta),$ we find that for $\delta>0$ NEC can be violated or satisfied depending upon $F(R)<0$ or $F(R)>0$ respectively. (Here $F(R)<0$ is not viable because then the matter density $\rho<0.$)

Further this being a power law $f(R)$ model, there are several cosmological implications of the model that have already been discussed corresponding to the $f(R)$ models in Solution 1 and 2. Although, there is an apparent similarity of the $f(R)$ models in Solutions 1 and 3 yet it may be worth mentioning that, they appear corresponding to two different wormhole metric, and their different parameter dependencies make them distinguishable and are independent (that is no parameter reduction makes them interchangeable). The range of applications of the above $f(R)$ models clearly demonstrate their astrophysical significance. Despite their variety of applications, the power-law models lacked a mathematical motivation. Our solution for the first time provides a mathematical foundation to these models. Solutions 1-4 shows that the power-law $f(R)$ models can be a generic solution for $f(R)$ with the eos (\ref{eos}). 

In FIG. 2 we have provided a graphical representation of the wormhole embedding function for Solutions 3 and 4 along with the $f(R)$ solutions (\ref{fr3}) and (\ref{fr4}) as a function of the curvature $R$ as obtained in (\ref{R3}) and (\ref{R4}). The sub-figure 2(a) clearly shows that the wormhole shape function in Solution 3 cannot become asymptotically flat. The parameters selected correspond to {\it Region IV} of the parameter space. That the wormhole shape function for Solution 4 is asymptotically flat is amply observed in FIG. 2(c).

\begin{figure}
\centering
\subfigure[]{\includegraphics[width=7cm,height=5cm]{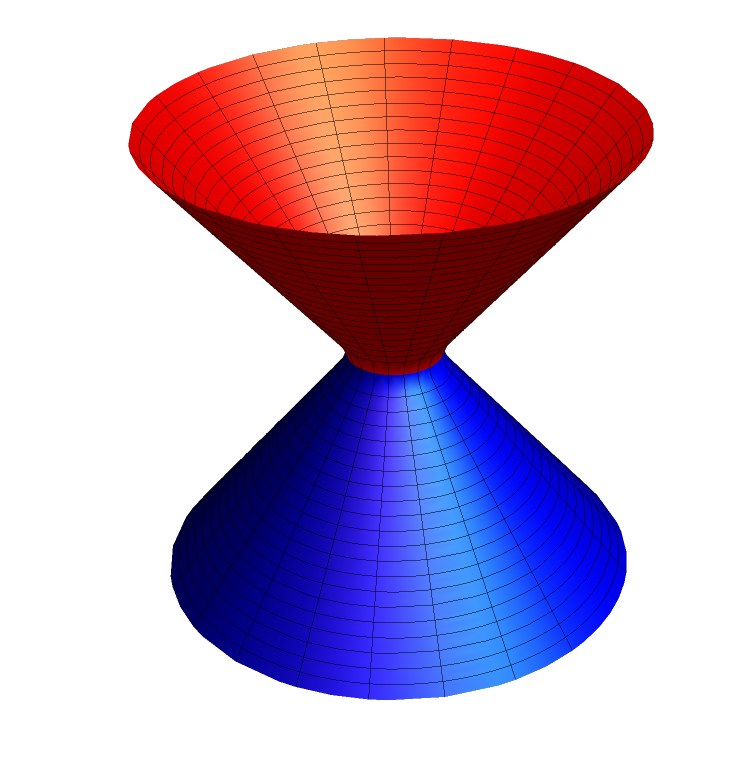}}\hspace{2cm}
\subfigure[]{\includegraphics[width=7cm,height=5cm]{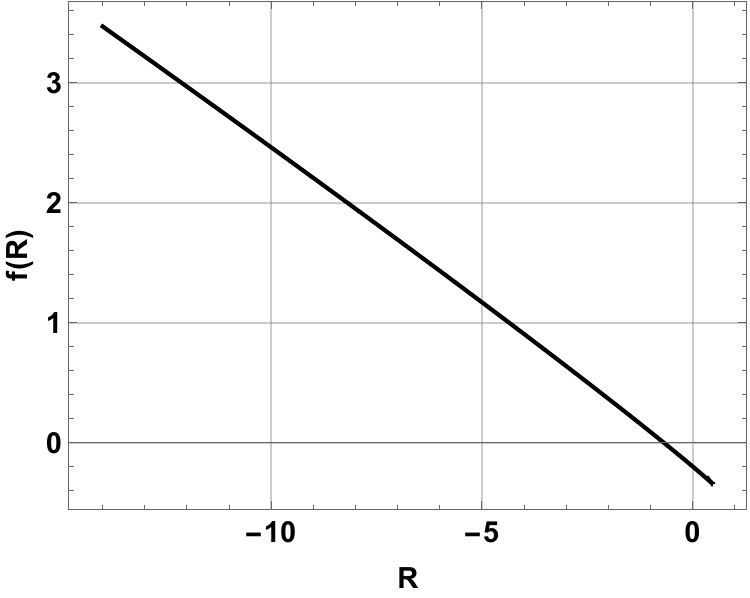}}\\
\subfigure[]{\includegraphics[width=7cm,height=5cm]{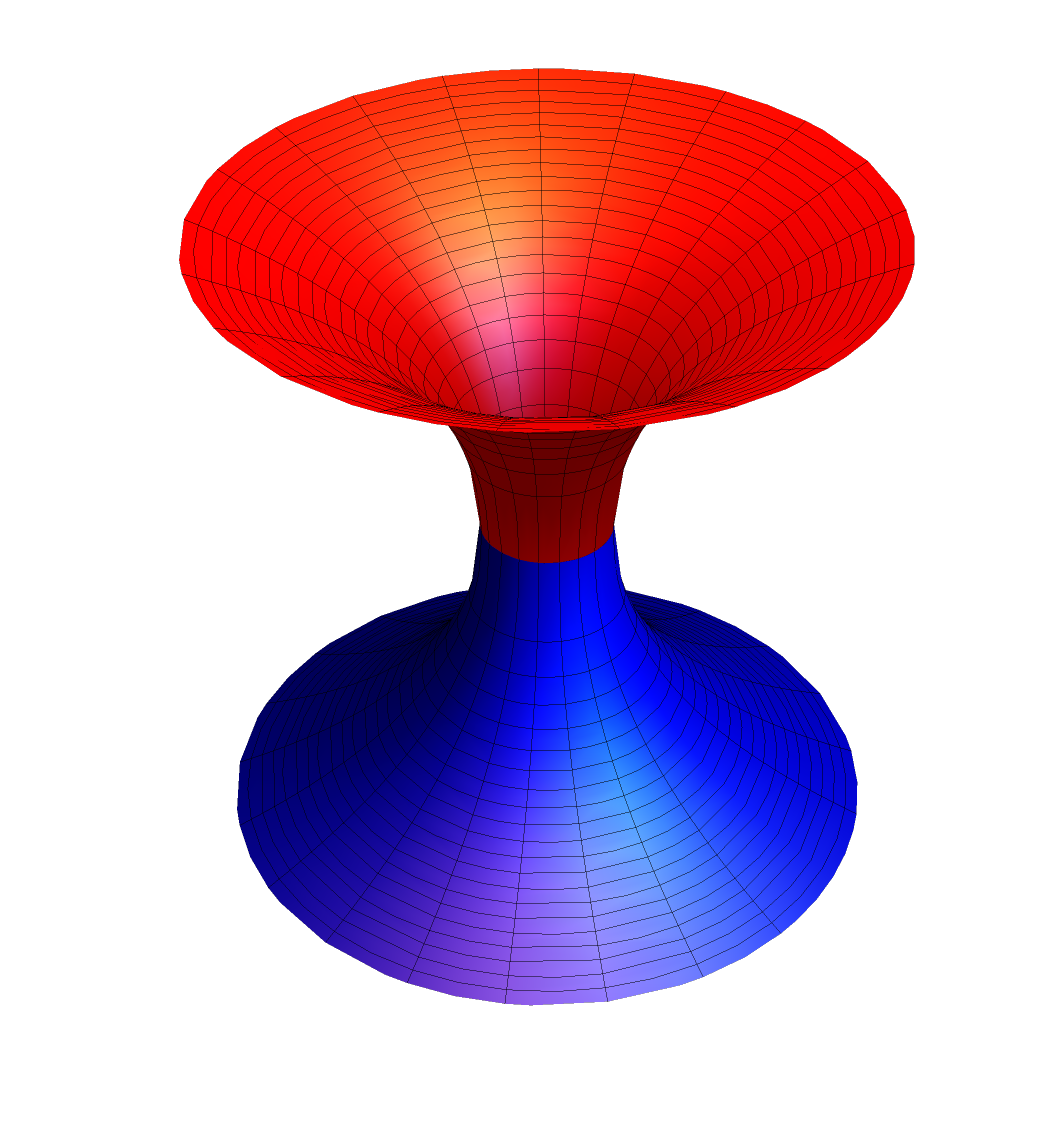}}\hspace{2cm}
\subfigure[]{\includegraphics[width=7cm,height=5cm]{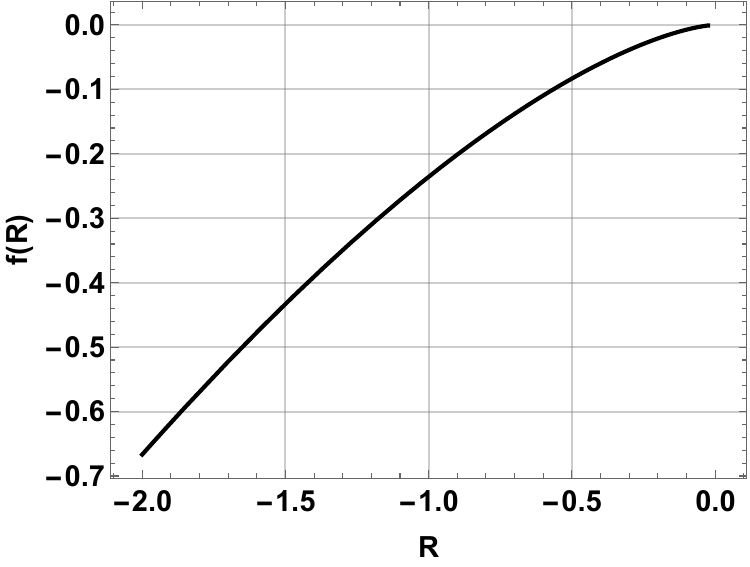}}
\caption{Figures (a) and (b) respectively depict the embedding diagram for the wormhole throat and the $f(R)$ as a function of the curvature $R$ corresponding to Solution 3. The figures have been obtained for $r_{0}=1,~\delta=-4,~\alpha=-1$ and $f_{1}=1.$ Figures (c) and (d) respectively show the embedding for the wormhole throat and the $f(R)$ solution as a function of $R$ corresponding to Solution 4. Both (c) and (d) have been constructed with parameters $r_{0}=1,~\delta=3,~f_{4}=-1.$}
\label{fig:Ftwo}
\end{figure}

\section{Energy Condition and existence of Ghost Field}

The energy conditions satisfied by the matter tensors in the vicinity of wormhole throat is a topic of great interest in wormhole studies. Geometry of the traversable wormhole affects the NECs of the matter tensors resulting in their violation. A bulk of wormhole literature therefore concentrates on mechanisms to obtain wormhole throat that does not require NEC violating matter. Gravity modification is most commonly used to address this issue \cite{gbll, fr, rs, rast}. In $f(R)$ gravity the the higher order curvature terms are interpreted as the effective matter-tensors. Specifically we can rewrite the equation (\ref{ffe}) as \cite{far,soti}:
\begin{equation}
G_{\mu\nu}=\frac{\kappa}{f_{R}(R)}\left[T^{m}_{\mu\nu}+T^{f}_{\mu\nu}\right]\label{fmatter}
\end{equation} 
where $G_{\mu\nu}$ is the usual Einstein tensor, $T^{m}_{\mu\nu}$ is the matter tensor, while $T^{f}_{\mu\nu}=\frac{1}{\kappa}\left[\frac{f(R)-Rf_{R}}{2}g_{\mu\nu}+\left(\nabla_{\mu}\nabla_{\nu}-g_{\mu\nu}\square \right)f_{R}\right]$ is the $f(R)$ curvature matter. For traversable wormhole geometry to be supported by an usual NEC satisfying matter tensors (ensuring positivity of $T^{m}_{\mu\nu}+T^{f}_{\mu\nu}$) will render $f_{R}<0.$  That is that the associated graviton field turns ghost \cite{bron}. Thus to get a traversable wormhole supported by NEC satisfying matter tensors in the presence of $f(R)$ gravity requires an effective trade-off for the ghost graviton field. 

We can rewrite the eos (\ref{eos}) as $\rho+p_{t}= -\frac{1}{\delta}(\rho+p_{r}).$ Clearly for a negative $\delta$ the NECs are satisfied for $\rho+p_{r}>0$ while for a positive $\delta$ the NEC is never satisfied. An analysis of the NECs in the Solutions 1-4 will show that it is possible to obtain wormhole, both with NEC violating and satisfying matter tensors and that in case of NEC satisfying matter tensors we always have an associated ghost $f(R)$.

Corresponding to Solution 1 we obtain (for $\kappa=1$):
\begin{equation}
\rho+p_{r}=-\left(\frac{F_{01}}{2r_{0}^{2}}\right)\left(1-\frac{\eta_{1}}{\xi_{1}}\right)\left(\frac{r_{0}}{r}\right)^{-\alpha}\left[\left\{2\alpha^{2}+\left(6+\frac{4}{\delta}\right)\alpha+\left(4+\frac{16}{\delta}\right)\right\}\left(1-\left(\frac{r_{0}}{r}\right)^{\xi_{1}}\right)+(\alpha+4)\xi_{1}\left(\frac{r_{0}}{r}\right)^{\xi_{1}}\right]\label{nec1}
\end{equation}
Since the NEC is satisfied only for $\delta<0$ and $\rho+p_{r}>0,$ we consider the parameter space for {\it Region I} only (it is sufficient to consider only $\rho+p_{r}$ as $\rho+p_{t}$ can be obtained as a scalar multiple of $\rho+p_{r}$). From equation (\ref{nec1}) we observe that $\rho+p_{r}>0$ for $F_{01}<0$ and $\alpha>-1,~\delta<\text{min}\left(-\frac{2\alpha+8}{\alpha^{2}+3\alpha+2},-(\alpha+6)\right).$ This means that for the parameter space of {\it Region I}, NEC is satisfied for whole of {\it Region I} except for a narrow strip $\delta\in(-5,-4)$ and $\alpha\in(-2,-1).$ Further $F_{01}<0$ necessarily implies that $F(R)<0$ in the parameter space of {\it Region I}. This renders the $f(R)$ gravity solution as ghost field. Thus in {\it Region I} the wormhole solution given by the metric representation (\ref{met1}) can be sustained by NEC satisfying matter fields provided the $f(R)$ solution (\ref{fr1}) is ghost. For {\it Region II} and {\it III} the NEC is not satisfied as $\delta >0.$ In this case we have $F_{01}>0$ which renders the associated $f(R)$ gravity non-ghost. 

Corresponding to Solution 3 we analyse the NEC by considering:
\begin{equation}
\rho+p_{r}=-\left(\frac{F_{03}}{2r_{0}^{2}}\right)\left(1-\frac{\eta_{3}}{\xi_{3}}\right)\left(\frac{r_{0}}{r}\right)^{-\alpha}\left[\left\{2\alpha^{2}+6\alpha+4\right\}\left(1-\left(\frac{r_{0}}{r}\right)^{\xi_{3}}\right)+(\alpha+4)\xi_{3}\left(\frac{r_{0}}{r}\right)^{\xi_{3}}\right]\label{nec3}
\end{equation}
Corresponding to the parameter space in {\it Region IV}, $\rho+p_{r}>0$ for $F_{03}<0$ and $\delta<-3,~\alpha\in[-1,-(\delta+4)).$ Again $F_{03}<0$ gives $F(r)<0.$ This means we have again obtained a wormhole solution with NEC satisfying matter in the presence of ghost $f(R)$ solution.

It may be noted that for wormhole solutions 1 and 3 not only is the NEC satisfied, but we observe that in {\it Region I} and {\it Region IV} respectively the weak energy conditions (WECs) and dominant energy conditions (DECs) are also satisfied. The WEC is obtained if in addition to NEC the matter tensors satisfy $\rho\geq 0.$ The DEC is obtained for $\rho-|p_{i}|\geq 0$ where $i=(r,t).$ 

Solutions 2 and 4 is independent of the parameter $\alpha$ and exits only for $\delta>0.$ Hence the NEC is violated corresponding to both these wormhole space times. 

Thus we can obtain the wormhole as given in Solution 1 and Solution 3 and {\it Region I} and {\it Region IV} in ghost $f(R)$ gravity with NEC satisfying matter tensors. While for all Solutions 1-4 with {\it Regions II, III, V} and $\delta>0$ the wormholes we have obtained will exist only in NEC violating matter. In FIG. 3 we have provided a graphical representation of the energy conditions corresponding to Solutions 1-4 in the vicinity of the wormhole throat. FIG. 3(a) and FIG. 3(c) was constructed corresponding to parameters in {\it Region I} and {\it Region IV} respectively. Clearly the WEC and DEC are both satisfied in these regions of the valid parameter space for $(\delta,\alpha).$ Further FIG. 3(b) and 3(d) clearly show the violation of the NEC in the vicinity of the throat for solutions 2 and 4. 

In FIG. 4 we present a graphical representation of the $f(R)$ solutions in 1 and 3 corresponding to {\it Region I-V}. Since in the parameter validity {\it Region I} and {\it Region IV} the $f(R)$ is ghost, we observe that for {\it Region I} and {\it Region IV} in FIG. 4(a) and (b) respectively the $f(R)$ is a decreasing function of $R$ while all non-ghost $f(R)$ vary in sync with $R.$ In fact Similar observations can be made in FIG. 1(b) and FIG. 2(b) where $f(R)$ function were decreasing wrt $R.$ This is because the parameters for which the figures were drawn corresponded to {\it Region I} and {\it Region IV} of the parameter space.

\begin{figure}
\centering
\subfigure[]{\includegraphics[width=7cm,height=5cm]{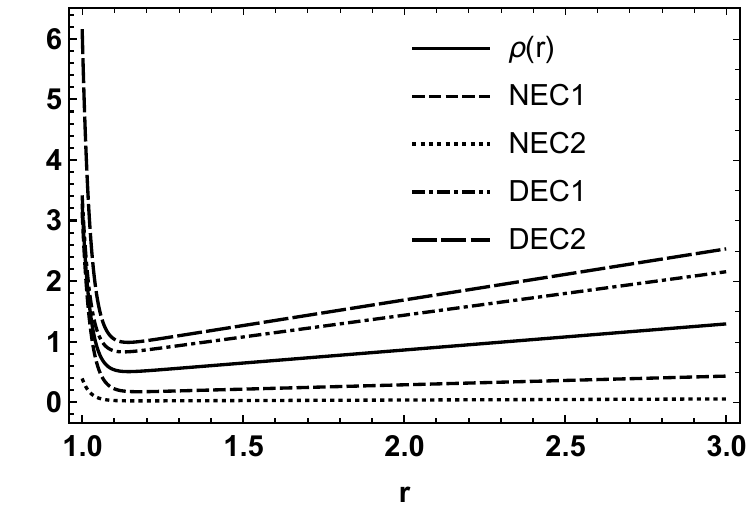}}\hspace{2cm}
\subfigure[]{\includegraphics[width=7cm,height=5cm]{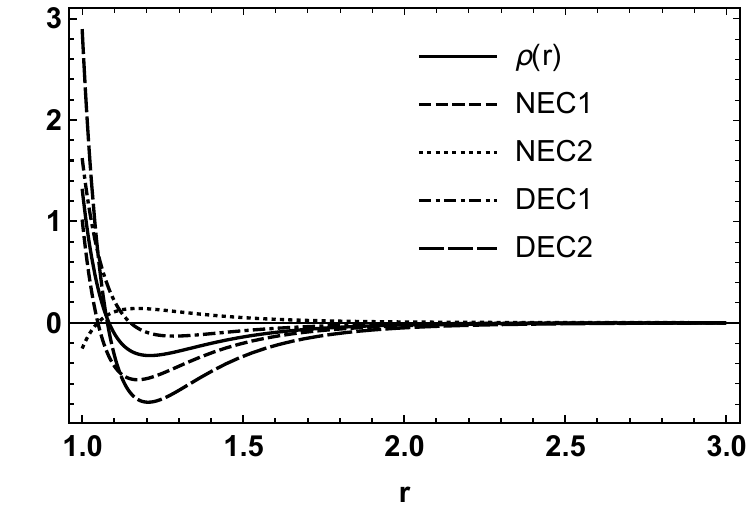}}\\
\subfigure[]{\includegraphics[width=7cm,height=5cm]{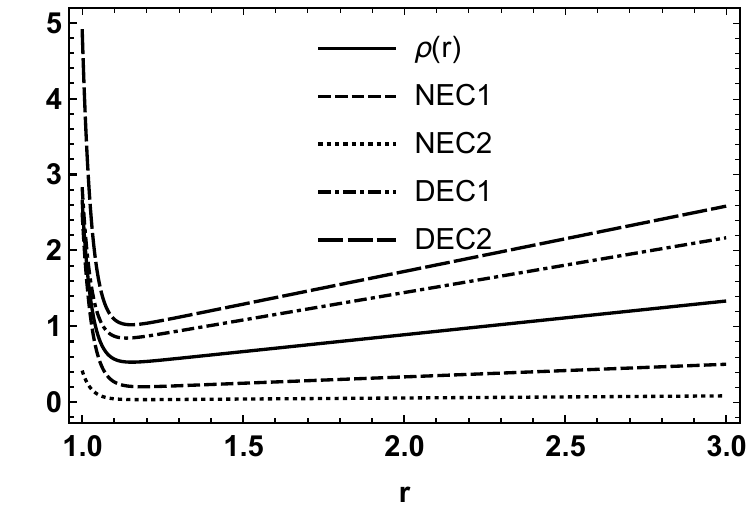}}\hspace{2cm}
\subfigure[]{\includegraphics[width=7cm,height=5cm]{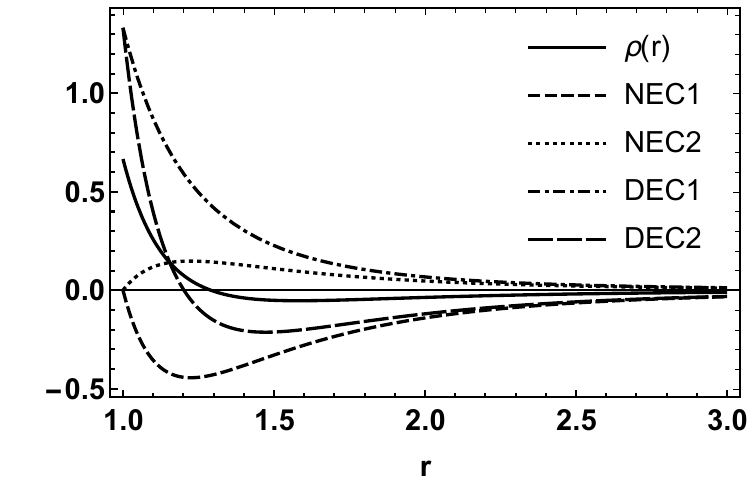}}
\caption{(a) The energy conditions corresponding to Solution 1, figure drawn corresponding to $r_{0}=1,~\delta=-8,~\alpha=1,~f_{1}=1.$ (b) The energy conditions corresponding to Solution 2, figure drawn corresponding to $r_{0}=1,~\delta=4,~f_{2}=1.$ (c) The energy conditions corresponding to Solution 3, figure drawn corresponding to $r_{0}=1,~\delta=-6,~\alpha=1,~f_{1}=1.$ (d) The energy conditions corresponding to Solution 4, figure drawn corresponding to $r_{0}=1,~\delta=3,~f_{4}=-1.$}
\label{fig:Fthree}
\end{figure}

\begin{figure}
\centering
\subfigure[]{\includegraphics[width=7cm,height=5cm]{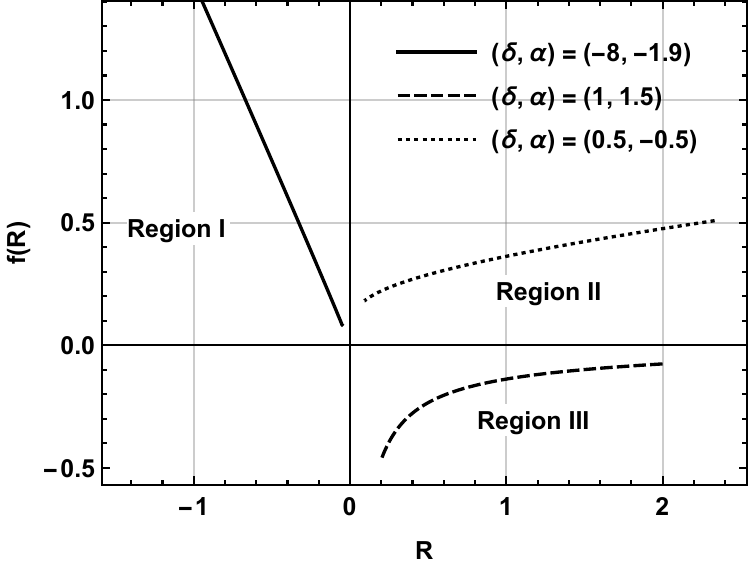}}\hspace{2cm}
\subfigure[]{\includegraphics[width=7cm,height=5cm]{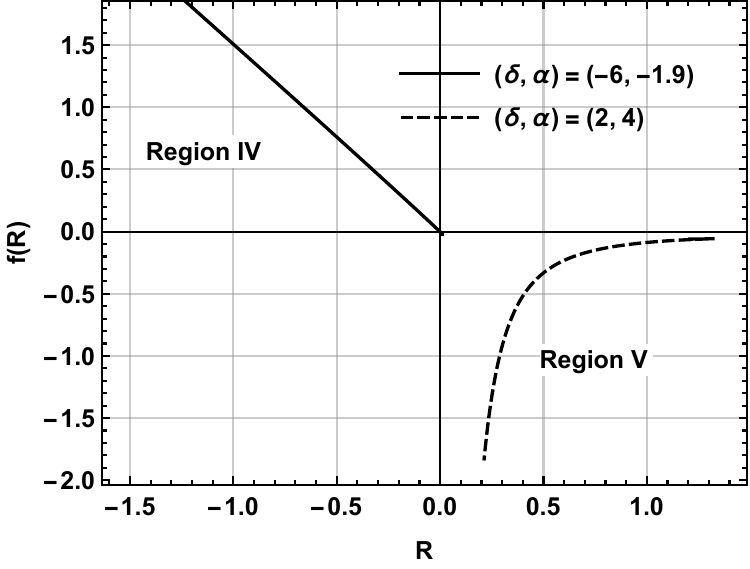}}
\caption{(a) Evolution of $f(R)$ wrt $R$ of Solution 1 in the parameter validity regions I-III. (b) Evolution of $f(R)$ wrt $R$ of Solution 3 in the parameter validity regions IV and V. Both figures were drawn for $r_{0}=1,~f_{1}=1.$ }
\label{fig:Ffour}
\end{figure}

\section{Equivalent Scalar-Tensor Representation using the Brans-Dicke Scalar Field}

The Brans-Dicke theory was proposed in 1961 by C. H. Brans and R. H. Dicke \cite{bd}. Metric $f(R)$ gravity theory was found equivalent to the BD scalar tensor theory for the particular value of the BD parameter $\omega=0$ \cite{chb}.  The representation of $f(R)$ in terms of the specific scalar-tensor theory have been useful in many aspects. For $f_{RR}(R)\neq 0$ the equivalence of $f(R)$ gravity with BD theory helped to show that the Cauchy problem was well-posed in $f(R)$ gravity \cite{cp}. $f(R)$ has found applications in wide range of studies from cosmology to astrophysics and its equivalence with BD gravity have been applied to establish its cosmological viability. 

Corresponding to a field $\sigma,$ the dynamically equivalent representation of the action (\ref{faction}) can be given as \cite{no,st1,st2}:
\begin{equation}
S_{\sigma}=\frac{1}{2\kappa}\int [f(\sigma)+f'(\sigma)(R-\sigma)]\sqrt{-g}d^{4}x.\label{naction}
\end{equation}
(In this section $'$ denotes derivative w.r.t the variable as per context of the function, that is $f'(\sigma)=\frac{df}{d\sigma}$ etc.) The variation of the above action wrt the field $\sigma$ gives $\sigma=R$ provided $f''(\sigma)\neq 0.$ This evidently gives the action (\ref{faction}) for $f(R).$ Redefining $\sigma=\sigma(\varphi)$ with $\varphi=f'(\sigma)$ and $W(\varphi)=\sigma f'(\sigma)-f(\sigma)$ and substituting in the equation (\ref{naction}) gives:
\begin{equation}
S_{\sigma}=\frac{1}{2\kappa}\int [R\varphi-W(\varphi)]\sqrt{-g}d^{4}x.\label{phiaction}
\end{equation} 
The BD action in Jordan frame is given as:
\begin{equation}
S_{BD}=\frac{1}{2\kappa}\int [R\varphi-\frac{\omega}{\varphi}g^{\mu\nu}\bigtriangledown_{\mu}\varphi\bigtriangledown_{\nu}\varphi-W(\varphi)]\sqrt{-g}d^{4}x.\label{bdaction}
\end{equation}
where $\omega$ is the BD parameter. Clearly the action (\ref{phiaction}) is a representation of the BD action in the Jordan frame (\ref{bdaction}) with the zero BD parameter $\omega$. Thus the metric $f(R)$ gravity action can be imagined as a representation of the BD gravity for a specific value of the BD parameter. In the presence of matter action $S_{M}$ the action (\ref{phiaction}) gives the field equation
\begin{equation}
\varphi G_{\mu\nu}+\left(g_{\mu\nu}\square-\triangledown_{\mu}\triangledown_{\nu}\right)\varphi+\frac{1}{2}g_{\mu\nu}W(\varphi)=\kappa T_{\mu\nu}\label{bdfe}
\end{equation}
This being same as the previously obtained $f(R)$ field equation (\ref{ffe}) for $\varphi=f_{R}=F(R)$ and $W(\varphi)$ as defined above. Clearly the $F(R)$ that we have obtained in Solutions 1-4 give suitable field representation $\varphi(r)$ for the $f(R)$ solutions (\ref{fr1}), (\ref{fr2}), (\ref{fr3}) and (\ref{fr4}). 
\begin{itemize}
\item Solution 1: $F(R(r))=F_{01}\left(\frac{r_{0}}{r}\right)^{-(\alpha+2)}$\\
This gives $W(\varphi)=\varphi_{01}\left[R_{1}\left(\frac{\alpha+2}{\alpha}\right)\left(\frac{\varphi}{\varphi_{01}}\right)^{\frac{\alpha}{\alpha+2}}+R_{2}\left(\frac{\alpha+2}{\alpha-\xi_{1}}\right)\left(\frac{\varphi}{\varphi_{01}}\right)^{\frac{\alpha-\xi_{1}}{\alpha+2}}\right]$ where $\varphi_{01}=\varphi(r_{0})$ corresponding to $F(R)$ Solution 1. 
\item Solution 2: $F(R(r))=F_{02}\left(\frac{r_{0}}{r}\right)^{\beta}$\\
This gives $W(\varphi)=\varphi_{02}\left[R_{3}\left(\frac{\beta}{\beta+2}\right)\left(\frac{\varphi}{\varphi_{02}}\right)^{\frac{\beta+2}{\beta}}+R_{4}\left(\frac{\beta}{\beta+2+\xi_{2}}\right)\left(\frac{\varphi}{\varphi_{02}}\right)^{\frac{\beta+2+\xi_{2}}{\beta}}\right]$ where $\varphi_{02}=\varphi(r_{0})$ for $F(R)$ Solution 2.
\item Solution 3: $F(R(r))=F_{03}\left(\frac{r_{0}}{r}\right)^{-(\alpha+2)}$\\
This gives $W(\varphi)=\varphi_{03}\left[R_{5}\left(\frac{\alpha+2}{\alpha}\right)\left(\frac{\varphi}{\varphi_{03}}\right)^{\frac{\alpha}{\alpha+2}}+R_{6}\left(\frac{\alpha+2}{\alpha-\xi_{3}}\right)\left(\frac{\varphi}{\varphi_{03}}\right)^{\frac{\alpha-\xi_{3}}{\alpha+2}}\right]$ where $\varphi_{03}=\varphi(r_{0})$ corresponding to $F(R)$ Solution 3. 
\item Solution 4: $F(R(r))=F_{04}\left(\frac{r_{0}}{r}\right)^{\delta-1}$\\
This gives $W(\varphi)=W_{0}\left(\frac{\varphi}{\varphi_{04}}\right)^{\frac{1+\delta+\xi_{4}}{\delta-1}}$ where $\varphi_{04}=\varphi(r_{0})$ for the above $F(R)$ in Solution 4 and $W_{04}=W(\varphi_{0})=R_{7}\varphi_{04}\left(\frac{\delta-1}{1+\delta+\xi_{4}}\right).$
\end{itemize}
Existence of wormholes in BD gravity was first studied in \cite{ag}. Later research \cite{bdwm} showed the existence of static wormholes (not necessarily traversable) in the Jordan frame for a very small range of the BD parameter. Research in \cite{lb} revealed that it is possible to find wormholes corresponding to a vanishing BD parameter. Here we have used the $f(R)$ gravity correspondence to obtain suitable BD scalar field and potential. The spherically symmetric, traversable wormholes obtained in solutions 1-4 can all exit in above constructed BD scalar fields with a potential field due to $f(R)$, in vanishing BD parameter without necessarily the field turning ghost. However as shown in the previous section, under suitable choice of parameter, the field has the possibility of becoming ghost whence the throat matter is non-exotic. This shows the existence of traversable wormhole in the vanishing BD parameter and in a non-ghost BD field. In a different context, it may be noted that similar scalar potentials have been previously associated with power law $f(R)$ for the dark energy description \cite{cap4} and with the inhomogeneous Fonarev solution \cite{vf}.

It may be noted that suitable Brans-Dicke scalar field can be obtained for the Palatini $f(R)$ gravity with $\omega=-\frac{3}{2}.$ This is however problematic because the ensuing dynamical scalar field equation loses its dynamical degree of freedom to become just an algebraic equation. As a result the scalar field $\phi$ effectively has no dynamical significance rendering the gravitational coupling $\frac{1}{\phi}$ ineffective. This also makes for an ill-posed Cauchy problem. However introducing a modification as hybrid metric-Palatini gravity \cite{harko} with gravitational coupling $\frac{1}{1+\varphi}$ renders the associated scalar field dynamical. This correction has found many useful applications in cosmology and astrophysics and has helped alleviate the shortcoming of Palatini and metric $f(R)$ gravity. In hybrid metric-Palatini gravity the field equation and the dynamical scalar field equation is modified as:
\begin{equation}
(1+\varphi) G_{\mu\nu}+\left(g_{\mu\nu}\square-\triangledown_{\mu}\triangledown_{\nu}\right)\varphi+\frac{1}{2}g_{\mu\nu}W(\varphi)+\frac{3}{2\varphi}\left[\triangledown_{\mu}\varphi\triangledown_{\nu}\varphi-\frac{1}{2}g_{\mu\nu}\triangledown_{\alpha}\varphi\triangledown^{\alpha}\varphi\right]=\kappa T_{\mu\nu}\label{fehy}
\end{equation}
\begin{equation}
-\frac{3}{\varphi}\square\varphi+\frac{3}{2\varphi^{2}}\triangledown_{\alpha}\varphi\triangledown^{\alpha}\varphi+2W(\varphi)-(1+\varphi)\frac{\partial W}{\partial\varphi}=\kappa T\label{sfhy}
\end{equation}
Using the wormhole solutions 1-4 (that is the red-shift functions $\phi$ and wormhole shape functions $b(r)$ obtained in Solutions 1-4) along with the corresponding BD scalar fields $\varphi$ as given above, in the equations (\ref{fehy}) and (\ref{sfhy}), we could obtain suitable potential $W(\varphi)$ and hence the corresponding NECs for the matter tensors $T_{\mu\nu}$ in hybrid metric-Palatini gravity. This exercise provides an independent test for the performance of the modelled solutions in hybrid gravity regime. The results are provided as below:
\begin{itemize}
\item Solution 1:\\ $W_{hbd}(\varphi)=\varphi_{01}\left[W_{01}\left(\frac{\alpha+2}{\alpha}\right)\left(\frac{\varphi}{\varphi_{01}}\right)^{\frac{\alpha}{\alpha+2}}+W_{11}\left(\frac{\alpha+2}{\alpha-\xi_{1}}\right)\left(\frac{\varphi}{\varphi_{01}}\right)^{\frac{\alpha-\xi_{1}}{\alpha+2}}\right]$ where\\$W_{01}=\frac{1}{2r_{0}^{2}\delta^{2}}\left[\frac{4\delta^{2}\eta_{1}}{\xi_{1}}-\left(1-\frac{\eta_{1}}{\xi_{1}}\right)\left(3(\alpha^{2}+6\alpha+8)\delta^{2}-4(3\alpha+8)\delta+16\right)\right]$ and\\$W_{11}=\frac{1}{2r_{0}^{2}\delta^{2}}\left(1-\frac{\eta_{1}}{\xi_{1}}\right)\left[\left(3\alpha^{2}+3\alpha(6-\xi_{1})+28-10\xi_{1}\right)\delta^{2}-4(3\alpha+8-\xi_{1})\delta+16\right].$ With this potential and remaining results of Solution 1, the NECs in hybrid metric-Palatini gravity are satisfied as follows:
\begin{itemize}
\item{\it Region I}: Complete NEC satisfied for $1+F_{01}\leq 0.$
\item{\it Region II}: Complete NEC satisfied for $0\leq 1+F_{01}\leq \frac{\alpha+2}{\alpha+4}.$
\item{\it Region III}: Complete NEC satisfied for $\alpha\in(-2,\infty)$ and  $0\leq 1+F_{01}\leq \frac{\alpha+2}{\alpha+4}.$
\end{itemize}
\item Solution 2:\\$W_{hbd}(\varphi)=\varphi_{02}\left[W_{02}\left(\frac{\beta}{\beta+2}\right)\left(\frac{\varphi}{\varphi_{02}}\right)^{\frac{\beta+2}{\beta}}+W_{12}\left(\frac{\beta}{\beta+2+\xi_{2}}\right)\left(\frac{\varphi}{\varphi_{02}}\right)^{\frac{\beta+2+\xi_{2}}{\beta}}\right]$ where\\$W_{02}=-\frac{1}{2r_{0}^{2}\delta^{2}}\left[3\beta(\beta-2)\delta^{2}+4(3\beta-2)\delta+16\right]$ and\\$W_{12}=\frac{1}{2r_{0}^{2}\delta^{2}}\left[(3\beta^{2}-3\beta(2-\xi_{2})+4-4\xi_{2})\delta^{2}+4(3\beta-2+\xi_{2})\delta+16\right].$  Here NEC is never satisfied at the throat.
\item Solution 3:\\ $W_{hbd}(\varphi)=\varphi_{03}\left[W_{03}\left(\frac{\alpha+2}{\alpha}\right)\left(\frac{\varphi}{\varphi_{03}}\right)^{\frac{\alpha}{\alpha+2}}+W_{13}\left(\frac{\alpha+2}{\alpha-\xi_{3}}\right)\left(\frac{\varphi}{\varphi_{03}}\right)^{\frac{\alpha-\xi_{3}}{\alpha+2}}\right]$ where\\$W_{03}=\frac{1}{2r_{0}^{2}}\left[\frac{4\eta_{3}}{\xi_{3}}-3\left(1-\frac{\eta_{3}}{\xi_{3}}\right)\left(\alpha^{2}+6\alpha+8\right)\right]$ and $W_{13}=\frac{1}{2r_{0}^{2}}\left(1-\frac{\eta_{3}}{\xi_{3}}\right)\left[3\alpha^{2}+3\alpha(6-\xi_{3})+28-10\xi_{3}\right].$ Corresponding to the results of Solution 3, the NECs in hybrid metric-Palatini gravity are satisfied as follows:
\begin{itemize}
\item{\it Region IV}: Complete NEC satisfied for $1+F_{03}\leq 0.$
\item{\it Region V}: Complete NEC satisfied for $\alpha\in(-2,\infty)$ and  $0\leq 1+F_{03}\leq \frac{\alpha+2}{\alpha+4}.$
\end{itemize}
\item Solution 4:\\$W_{hbd}(\varphi)=\varphi_{04}\left[W_{04}\left(\frac{\varphi}{\varphi_{04}}\right)^{\frac{\delta+1}{\delta-1}}+W_{14}\left(\frac{\delta-1}{1+\delta+\xi_{4}}\right)\left(\frac{\varphi}{\varphi_{04}}\right)^{\frac{1+\delta+\xi_{4}}{\delta-1}}\right]$ where $W_{04}=\frac{3}{2r_{0}^{2}}(3-\delta)(\delta-1)^{2}(1+\delta)^{-1}$ and $W_{14}=\frac{1}{2r_{0}^{2}}\left[3\delta^{2}+3(\xi_{4}-4)\delta+13-7\xi_{4}\right].$ This satisfies the complete NEC for $\delta\in(0,1)$ and $0\leq 1+F_{04}\leq \left(\frac{\delta-1}{\delta-3}\right).$
\end{itemize}
Wormholes partially resembling Solution 3 and Solution 4 were previously studied in hybrid metric-Palatini gravity by \cite{cap6}. Solution II of \cite{cap6} was characterized with throat shape function and scalar field similar to the Solution 4 in the current study. Our results echo their results where wormhole with matter tensors satisfying NEC was obtained in hybrid metric-Palatini gravity. The stability of this new fluid can be again analysed using the squared adiabatic sound speed $c_{s}^{2}$ as shown in Solution 4. Here however the fluid is not restricted by any equation of state and hence as in \cite{cap3} we isotropize the pressure as $p_{avg}=\frac{1}{3}(p_{r}+2p_{t})$ and evaluate $\frac{\partial p}{\partial \rho}=\frac{\frac{dp}{dr}}{\frac{d\rho}{dr}}.$ We find that the fluid is causal with $0\leq c_{s}^{2}<1$ for the ranges of parameters $\delta\in(0,1)$ and $\max\left\{0,\frac{(\delta-1)(5\delta^{2}+6\delta-9)}{5\delta^{3}-7\delta^{2}-27\delta+45}\right\}<(1+F_{04})\leq\min\left\{\left(\frac{\delta-1}{\delta-3}\right),\frac{(\delta-1)(2\delta^{2}+6\delta-9)}{2(\delta^{3}+\delta^{2}-9\delta+9)}\right\}$. Thus wormhole Solution 4 with the associated scalar field in the hybrid metric-Palatini gravity gives stable traversable wormhole with standard NEC satisfying matter.
\begin{figure}
\centering
\subfigure[]{\includegraphics[width=7cm,height=5cm]{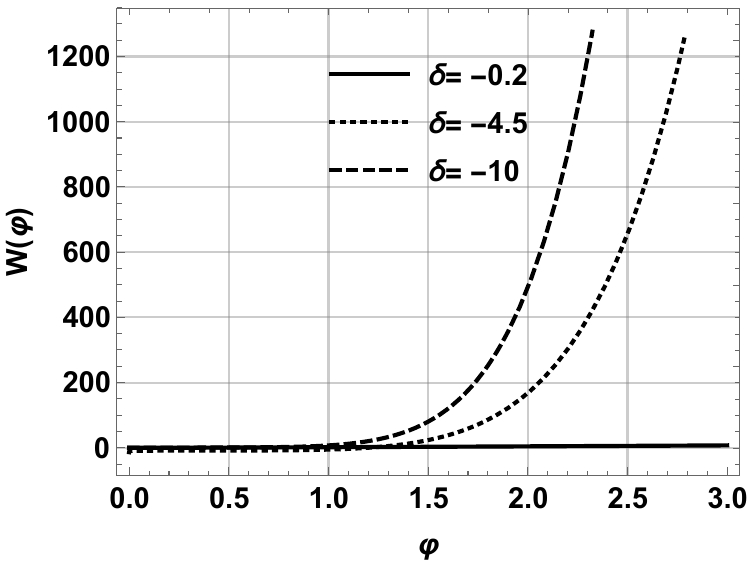}}\hspace{2cm}
\subfigure[]{\includegraphics[width=7cm,height=5cm]{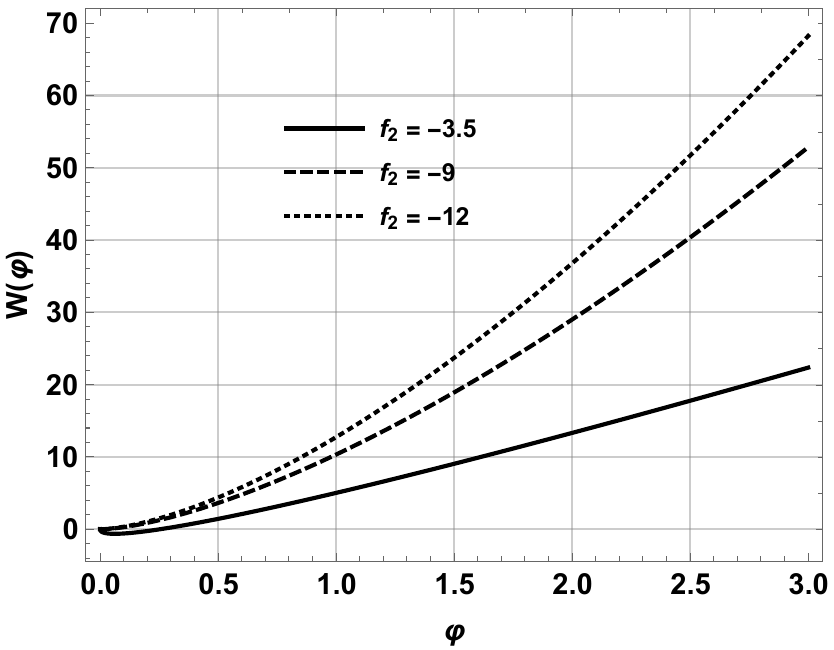}}\\
\subfigure[]{\includegraphics[width=7cm,height=5cm]{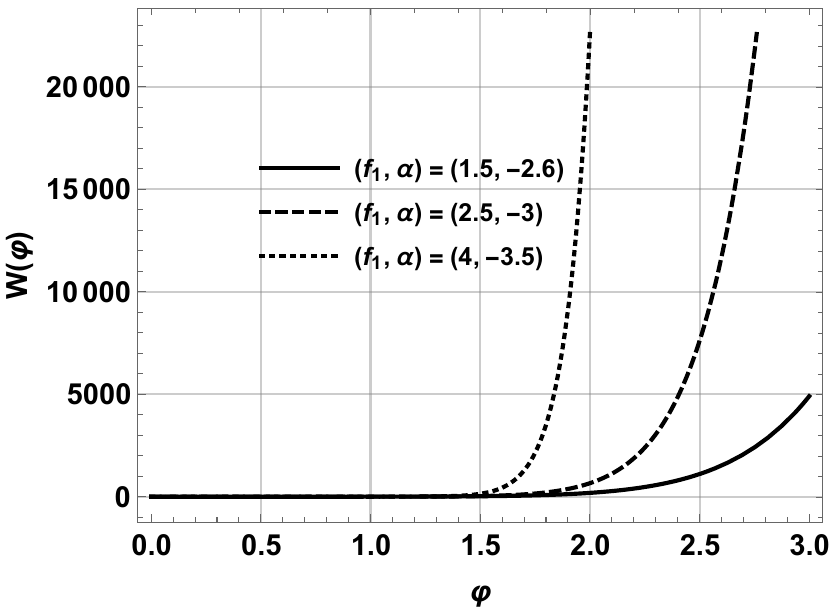}}\hspace{2cm}
\subfigure[]{\includegraphics[width=7cm,height=5cm]{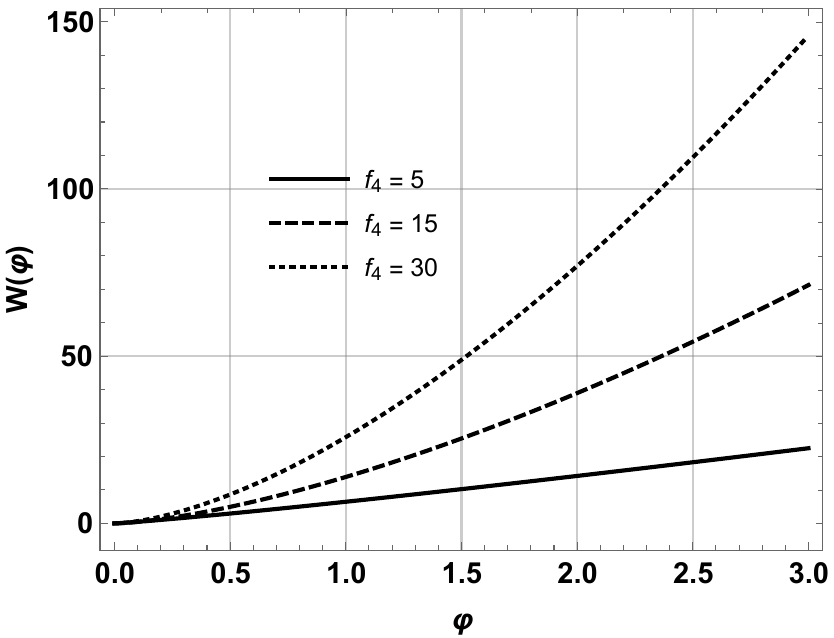}}
\caption{Figures (a)-(d) give the evolution of the Potential function $W(\phi)$ wrt the BD scalar $\phi$ corresponding to the Solutions 1-4 respectively. The parameters were chosen such that they satisfy the cosmological viability of the $f(R)$ models. Table 1 provides the parameter inclusions for cosmological viability of $f(R)$ models 1-4.}
\label{fig:Ffive}
\end{figure}

\subsection{Cosmological Significance of Non-Ghost $f(R)$ models}

Cosmological models of the universe prescribe exotic matter make-up like dark energy and dark matter, resulting in a present-time accelerated expansion of the universe. Gravity modification has been considered as a viable alternative to the exotic dark energy. $f(R)$ gravity has been proposed as an alternative to dark energy models \cite{defr} and has been used to propose a unified cosmic evolution of the universe starting from inflation to late time accelerating expansion of the universe \cite{unfr}. However such models have been criticised for being unstable and inconsistent with local gravity tests \cite{stfr}. Correspondence of $f(R)$ gravity with BD gravity for $\omega=0$ was also considered its limitation as a viable cosmological model \cite{chiba}. Such limitations were overcome by adopting the Chameleon mechanism for testing $f(R)$ in local high curvature regimes. Here we present some conditions for cosmologically viable $f(R)$ gravity models \cite{vifr}.
\begin{enumerate}
\item In order to avoid the formation of ghost scalar fields it is required that $f_{R}=F(R)>0.$

\item $f_{RR}>0$ is essential to prevent the scalaron field from turning tachyonic. This also ensures the existence of a stable matter dominated epoch before a late time accelerated universe. (It may be noted that the conversion of $f(R)$ gravity into the scalar-tensor BD gravity is possible only for $f_{RR}\neq 0$).

\item For successful big bang nucleosynthesis and Cosmic Microwave Background it is required that $f(R)\rightarrow R$ as $R\rightarrow \infty.$ 

\item Solar system and galactic survey tests assumes that $f(R)$ will predict galaxy formation similar to GR. This means, in a model of $R+f(R),~|f_{R}|$ is restricted to a very small value of $10^{-6}.$ Although such bounds might not be strict, yet a small value of $f_{R}$ in the current epoch is considered a success. For our model this means that in the current epoch $|f_{R}|=F(R)\simeq 1.$ 
\end{enumerate}
We can test for the cosmological viability of the $f(R)$ models provided in the Solutions 1-4 by their compliance to the above four conditions. Since our $f(R)$ models are $(\alpha,\delta)$ parameter dependent, we can put suitable parameter restrictions to ensure their cosmological acceptance. In Table 1 we provide the details of the parameter restrictions that are required for the above models to be cosmologically justified. The parameters in Table 1 were all calculated for $\delta<0$ for which the energy conditions are satisfied. (It may be noted that the results in Table 1 hold for $R\geq R(r_{0})$ where $R(r_{0})$ is the curvature at the wormhole throat. In the cosmological aspect, it may be considered as the curvature at the present epoch).

FIG. 5 provide a graphical visualization of the $f(R)$ potential $W(\phi)$ corresponding to the BD scalar $\phi$ and for parameters in the region of cosmological viability of the $f(R)$ solutions, as given by Table. 1. In all these models $\delta<0$ such that the NEC holds in the cosmological scenario.

\renewcommand{\arraystretch}{2.5} 
\begin{table}
\setlength\tabcolsep{4pt}
\setcellgapes{3pt}
\makegapedcells
    \begin{tabularx}{\textwidth}{c|X|X|c|l}
       \hhline{=====}
\thead{$f(R)$ Model} & \thead{$\delta$} & \thead{$\alpha$} & \thead{$f_{i},~i=1,2,4$}&\thead{$r_{0}$}\\
       \hhline{=====}
Solution 1 &$(-0.334,0)\cup(-\alpha,-2)$&$\alpha=-\frac{\delta^{2}+5\delta+2}{\delta}$&$f_{1}=\frac{2\delta^{2}+6\delta+4}{\delta}$&1\\
        \hline
Solution 2
&$\frac{f_{2}-1-\sqrt{(f_{2}-1)^{2}-8}}{2}<0$&$--$&$f_{2}<-3.13$&1\\
        \hline
Solution 3&$\frac{f_{1}-(\alpha+1)(\alpha+2)}{\alpha+2}<0$&$-4<\alpha<-2.5$&$(\alpha+2)(\alpha+1)<f_{1}<-3(\alpha+2)$&1\\
\hline
Solution 4&$1-f_{4}<0$&$--$&$f_{4}>4$&1\\
\hhline{=====}
\end{tabularx}
\caption[One]{The valid parameter restrictions for cosmological significance of $f(R)$ models}
\end{table}

\section{Light Rings of the Wormhole Models and associated Complexity Factor}

Features like light rings around super massive regions of space time have become important observables for predicting and testing GR and elemental physics. In the recent past the black hole images observed by the Event Horizon Telescope have paved the way for not just testing black hole physics, but also exotic objects like wormhole, naked singularity gravastars and may be even higher order gravity theories like $f(R).$ In this context, we explore the existence of light rings and photon spheres of the wormhole solutions 1-4. It has been shown previously in \cite{raj, snsb3} that corresponding to the wormhole metric (\ref{metric}) a light ring exists if 
\begin{equation}
\frac{2e^{2\phi}}{r^{2}}\sqrt{1-\frac{b}{r}}\left(\phi'(r)-\frac{1}{r}\right)=0\label{lr1}
\end{equation}
and that, a light rings can be a photon sphere (unstable light ring) if
\begin{equation}
2\left(1-\frac{b(r)}{r}\right)\frac{e^{2\phi}}{r^{2}}\left[2(\phi')^{2}+\phi''-\frac{4}{r}\phi'+\frac{3}{r^{2}}\right]+\frac{e^{2\phi}}{r^{3}}\left(\frac{b(r)}{r}-b'(r)\right)\left(\phi'-\frac{1}{r}\right)<0.\label{lr2}
\end{equation}
The equation (\ref{lr1}) show, that for a wormhole, a light ring always exists at the throat $r=r_{0}$ since $b(r_{0})=r_{0}$. Further this will be a photon sphere if $\phi'(r_{0})<\frac{1}{r_{0}}$ ($b'(r_{0})<1$ due to throat flare out condition). Other light rings might also exist at some location outside the throat ($r>r_{0}$), and hence a photon sphere depending on whether the conditions (\ref{lr1}) and (\ref{lr2}) holds for some other value of $r (\neq r_{0}).$ Analysing the four wormhole solutions 1-4, using equation (\ref{lr1}) we observe that, all four wormhole will have only one light ring at the throat. Using the equation (\ref{lr2}) we find that the light rings corresponding to all four wormholes will also be a photon sphere for valid ranges of the parameters $(\alpha,\delta)$.
In a recent paper \cite{snsb3} the photon spheres of a general wormhole in GR were connected to their complexity factor. Complexity factor \cite{her1} is a scalar quantity that can map the curvature of the space time with the corresponding matter make up. Space-times with zero complexity are considered to ``simple" space-times that are equivalent to a homogeneous and isotropic universe like the Minkowskii space-time. A space-time characterized by non-zero complexity will then signify its departure from homogeneity and isotropy. Thus the complexity of a geometry tells us about the inhomogeneity and anisotropy of the corresponding body of fluid. In fact general wormholes are generically complex objects that can never have zero complexity \cite{snsb1}. However wormholes with zero tidal force, characterized by $\phi'(r)=0$ will always have zero complexity. This is because, in such wormholes the absence of tidal forces ensures geodesic flow of the fluid, making the system stable with zero-complexity \cite{snsb2,her2}. Thus in GR our wormhole solutions 3 and 4 will always have zero complexity due to the absence of tidal forces ($\phi'(r)=0$). Here we have defined our wormholes in the presence of $f(R)$ gravity. In the presence of $f(R)$ gravity and wormhole metric (\ref{metric}) the complexity factor can be evaluated as:
\begin{equation}
Y_{TF}=-\frac{\kappa}{2F}\Pi+E+\frac{H}{2F}\label{ytf}
\end{equation}
where
\begin{align}
\Pi&=p_{r}-p_{t}\\
E&=\frac{1}{2}\left(1-\frac{b(r)}{r}\right)\left(\phi''+(\phi')^{2}-\frac{\phi'}{r}\right)+\left(\frac{rb'(r)-b(r)}{4r^{2}}\right)\left(\frac{1}{r}-\phi'\right)-\frac{b(r)}{2r^{3}}\\
H&=\left(1-\frac{b(r)}{r}\right)\left(-F''+\frac{F'}{r}\right)+F'\left(\frac{rb'(r)-b(r)}{2r^{2}}\right)
\end{align}
Substituting the values of $p_{r}$ and $p_{t}$ from the field equations (\ref{fe2}) and (\ref{fe3}) we get:
\begin{equation}
Y_{TF}=\left(1-\frac{b(r)}{r}\right)\left(\phi''+(\phi')^{2}-\frac{\phi'}{r}\right)-\left(\frac{rb'(r)-b(r)}{2r^{2}}\right)\phi'(r)\label{ytffin}
\end{equation}
which is basically the complexity factor in GR in terms of the metric coefficients. Thus we observe that complexity of a wormhole remains unaltered in $f(R)$ modification of gravity. This is expected because the geometric properties of the wormhole space-time remains unaltered in the presence of $f(R)$ gravity where $f(R)$ modification is mediated via adjustment in the matter tensors as has been explained in equation (\ref{fmatter}). Clearly one can observe from equation (\ref{ytffin}) that complexity is zero for any wormhole space-time with $\phi'(r)=0.$ Also, at the throat $r=r_{0}$ the complexity reduces to:
\begin{equation}
Y_{TF}(r_{0})=-\left(\frac{b'(r_{0})-1}{2r_{0}}\right)\phi'(r_{0})\label{ytf0}
\end{equation}
Since the shape functions $b(r)$ and the red-shift function $\phi(r)$ for wormhole solutions 1  and 2 have dependence on the parameters $(\alpha,\delta)$ the complexity will also depend upon them. In \cite{snsb3} a mechanism of identifying the existence of photon spheres based on complexity at the location of the light ring was suggested. Using (\ref{ytf0}) we can evaluate the complexity of the wormhole solutions 1 and 2 at the throat. Following the results of \cite{snsb3} we predict the existence of photon sphere at the throat based on the evaluated value of the complexity.  
\begin{itemize}
\item Solution 1: For solution 1 we obtain the complexity $Y_{TF}(r_{0})\begin{cases}<0,~\delta>0\\>0,~\delta<0\end{cases}.$ 

Following \cite{snsb3} we conclude that light ring at $r_{0}$ will be a photon sphere for $Y_{TF}<0,$ that is for $\delta>0.$ While for $Y_{TF}>0$ the light ring will be a photon sphere if it satisfies $Y_{TF}(r_{0})+\frac{b'(r_{0})-1}{2r_{0}^{2}}<0.$ We observe that this condition is satisfied for $\delta<-2.$ Since wormhole solution 1 exists for $\delta<-4$ we conclude that, again the light ring in solution 1 is a photon sphere.  
\item Solution 2: Here again $Y_{TF}(r_{0})\begin{cases}<0,~\delta>0\\>0,~\delta<0\end{cases}.$ Since the wormhole in solution 2 exists for only $\delta>\frac{2}{3}$ the light ring at the throat is always a photon sphere.
\item Solutions 3 and 4: Here the red-shift function $e^{\phi(r)}$ is a constant, hence the complexity factor is always zero for such wormholes. And such wormholes will always have a photon sphere at the throat. 
\end{itemize}
Thus analysing the complexity of wormhole space time we could independently corroborate the existence of the photon sphere at the throat corresponding to valid parameter range for all solutions 1-4. FIG. 6 is a graphical presentation of the photon spheres at the throat location for wormholes 1-4.

\begin{figure}
\centering
\subfigure[]{\includegraphics[width=4cm,height=4cm]{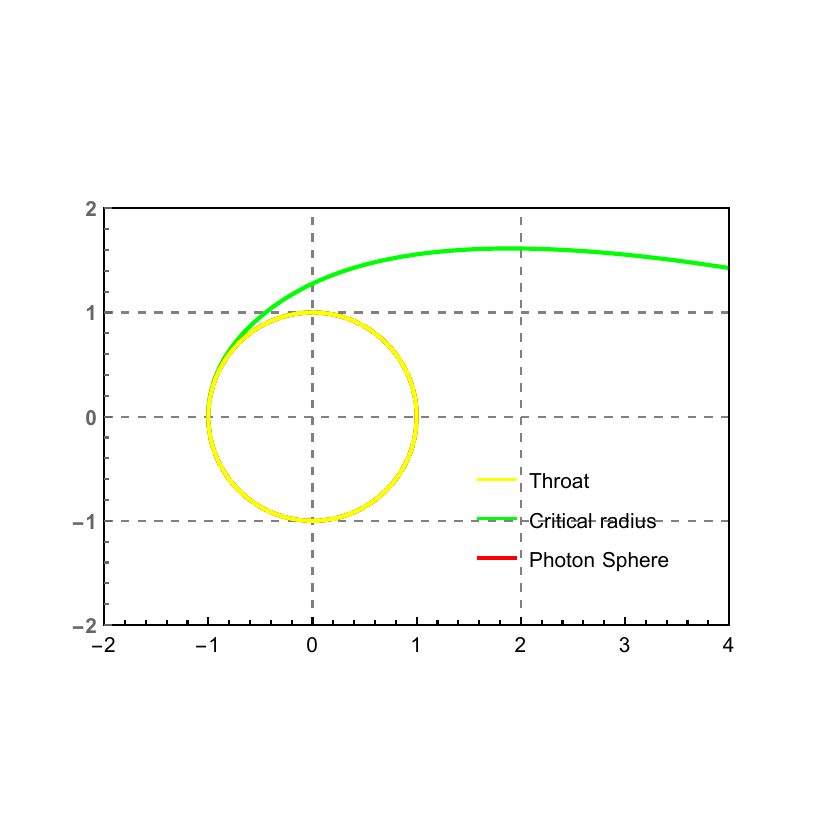}}\quad
\subfigure[]{\includegraphics[width=4cm,height=4cm]{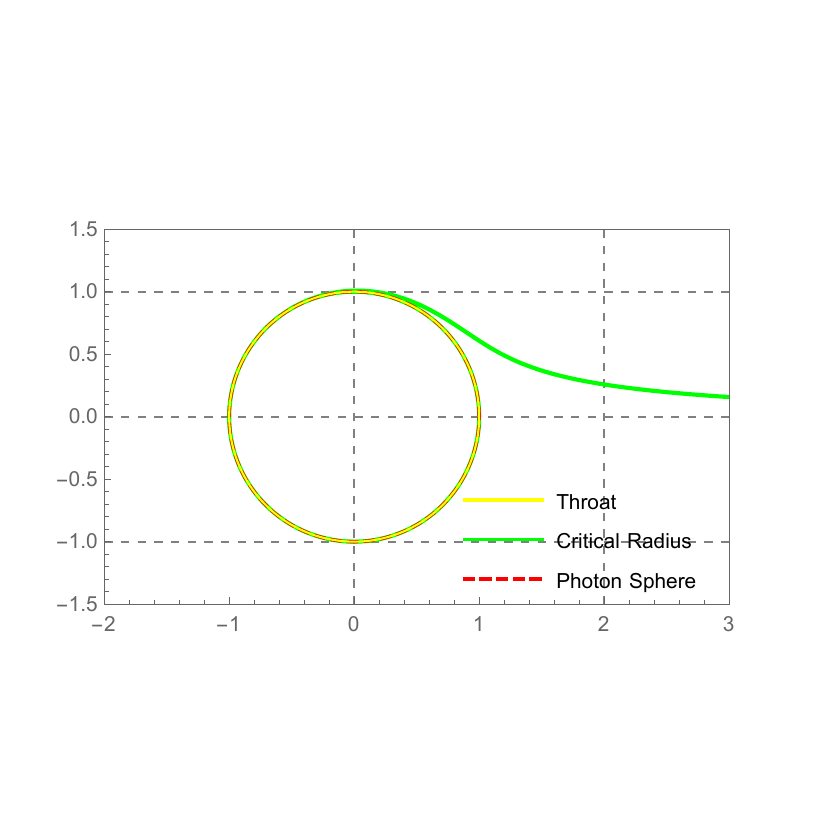}}\quad
\subfigure[]{\includegraphics[width=4cm,height=4cm]{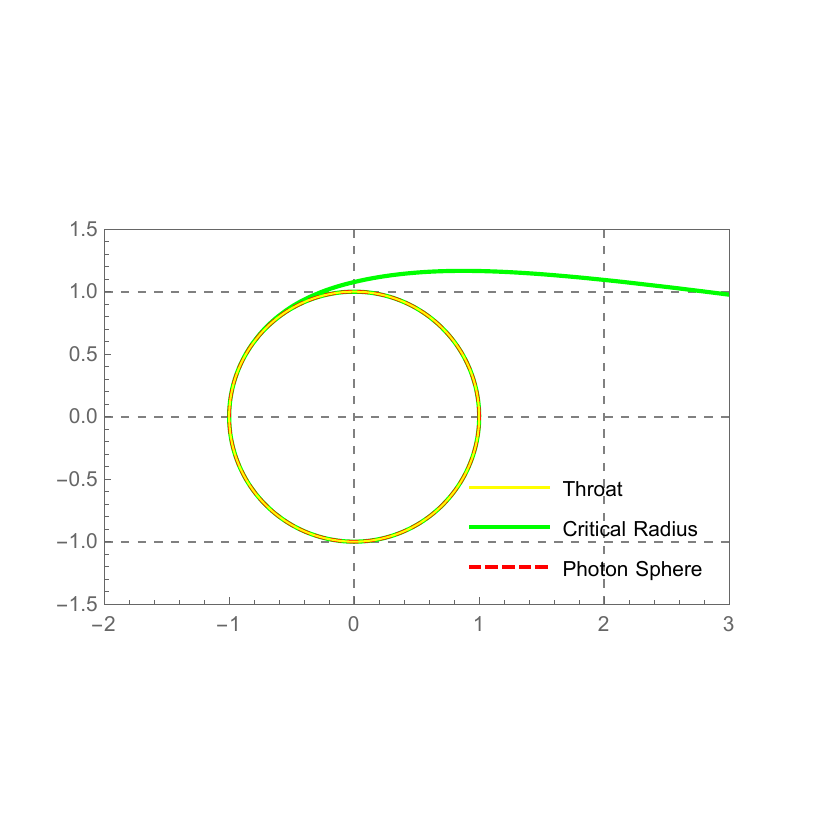}}\quad
\subfigure[]{\includegraphics[width=4cm,height=4cm]{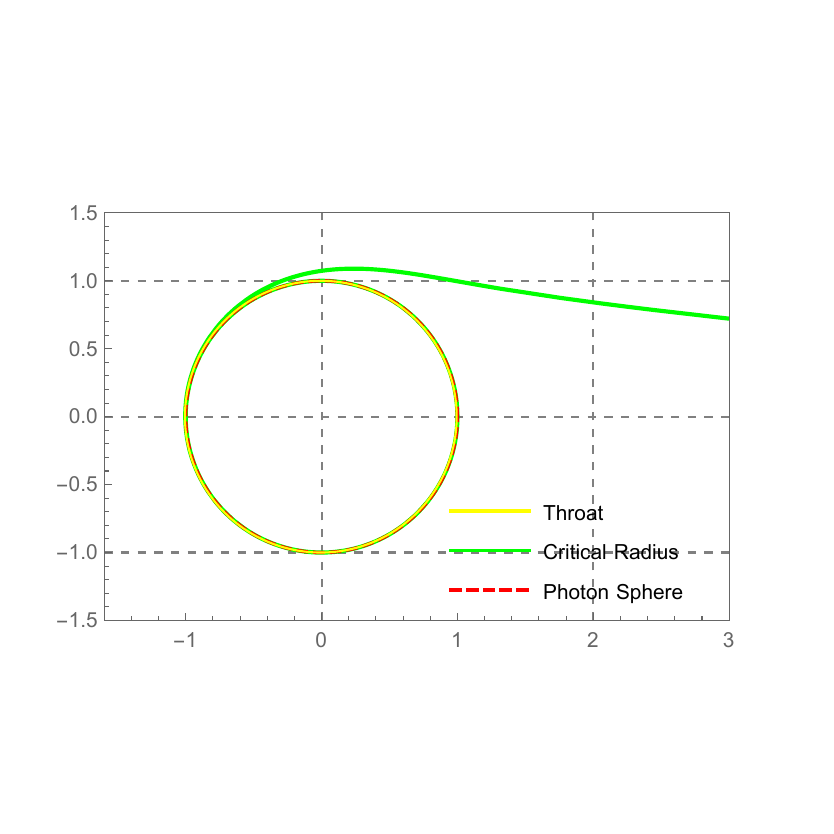}}
\caption{The figures (a)-(d) is a visual representation of the formation of photon sphere at the location of the throat in wormholes 1-4 respectively. The critical radius corresponds to the geodesics that can form a light ring. The throat is chosen at $r_{0}=1.$ }
\label{fig:Fsix}
\end{figure}

\section{Conclusion}

Starting from a novel phenomenological linear eos for the effective matter component in the inhomogeneous Morris-Thorne wormhole metric, we envisaged a plan for obtaining independent $f(R)$ and wormhole models. Here we used mathematical methods to independently construct four novel configurations of $f(R)$ and wormhole. The shape functions for wormholes in Solutions 1 and 3 have a solid angle deficit, as a result they are finite sized and not asymptotically flat. In GR such wormhole geometries have been previously explored in the context of isotropic pressure. Shape functions for wormhole in Solutions 2 and 4 are spherically complete with no solid angle incompleteness. They are asymptotically extendible. However the red-shift function of wormhole in Solution 2 has an event horizon as $r\rightarrow\infty.$ This being an undesirable feature of wormhole geometry, the wormhole is engineered so that it is finite sized and is not extendible upto infinity. In figures 1(a), 1(c), 2(a) and 2(c) we have shown the three dimensional embedding of the wormhole throats in Solutions 1-4 respectively. Using the figures we can amply visualise the asymptotic flatness of Solutions 2 and 4 as opposed to those of Solutions 1 and 3. It is noteworthy that wormhole Solution 4 is an example of the simplest traversable wormhole solution that is common and extensively used in the literature. Further, on account of the novelty of the assumed eos (\ref{eos}), the isotropization of the matter with averaged pressure leads to stiff matter supported wormhole, which corresponds to a non-ghost $f(R)$. Corresponding to the $f(R)$ solution 4 we emphasise that such solution already has been described in the literature and finds wide range of cosmological and astrophysical applications. It is interesting that we could suggest a  mathematical motivation in the background of an underlying physically viable  scenario for the wormhole and $f(R)$ model of Solution 4. Remaining $f(R)$ solutions are all implicit power law in $R(r)$ and such $f(R)$ models have applications in describing the cosmology of the universe. Explicit dependence of the $f(R)$ models wrt the curvature $R$ was graphically depicted in Figures 1(b), 1(d), 2(b), 2(d) and in FIG. 4. 

Another unique aspect of our study was, we did not impose the NEC on the $f(R)$ gravity solutions. Our models could be supported by both NEC satisfying as well as violating matter tensors. For the wormholes to exist in non-exotic matter, we had to compromise the $f(R)$ model which became ghost. Thus our motivation was to independently obtain $f(R)$ and wormhole solutions and not just those that would support wormholes without exotic matter. Figure 3 clearly show that wormholes in solutions 1 and 3 do not just satisfy the NEC they also satisfy the WEC and DEC. 

Using the equivalence of metric $f(R)$ gravity with BD scalar-tensor gravity, we provided the scalar field and scalar potential perspective of our models. It is clear that the wormholes modelled in our solutions can be supported by a scalar field with power-law potential. The robustness of the scalar field in the perspective of the wormhole solutions were additionally tested in the hybrid metric-Palatini gravity. Our results in hybrid gravity showed that solutions 1, 3 and 4 could be supported by NEC satisfying standard fluid in the presence of hybrid gravity modification. The fluid in Solution 4 was also found causal and completely stable against perturbations in the parameter space where NEC is satisfied. We also studied the cosmological fitness of the $f(R)$ models, irrespective of the wormholes and viable parameter space was obtained, where the $f(R)$ models in Solutions 1-4 could be applied to describe the cosmological evolution of the universe.

Observational astrophysics have acquired momentum due successful imaging of massive gravity regions in our space-time by the Event Horizon Telescope. The basis of such observation is the identification of photon spheres and shadows of objects like black hole, wormhole, naked singularity. Therefore, as an independent astrophysical study of the wormhole models, we analysed them for the existence of light rings and photon spheres. We further applied the results of \cite{snsb3} to show how the complexity of the wormholes at a certain radius in the vicinity of the throat could be used as a useful tool in detecting the photon sphere of the wormhole.

Corresponding to the eos (\ref{eos}), the above study prescribes a mathematical approach to obtain generic solutions for traversable wormholes and background $f(R)$ gravity. The models for both $f(R)$ and wormhole were demonstrated to have independent cosmological and astrophysical significance. As a future study, the $f(R)$ models could be applied as an alternative dark energy model in the expanding universe, while the wormhole solutions could be applied in the context of other gravity theories and in the context of other astrophysical probes.

\section{ Acknowledgements:}

SN acknowledges UGC, Government of India, for financial assistance through junior research fellowship (NTA ref.no. 231610097492). SB acknowledges IUCAA, Pune, India, where the project was initiated. SB acknowledges the support from the Department of Science and Technology (DST), Govt. of India under the Scheme
“Fund for Improvement of S\&T Infrastructure (FIST)”
(File No. SR/FST/MS-I/2019/41).


\begin{thebibliography}{}

%f(R) History
\bibitem{hw}H. Weyl, Annalen der Physick, {\bf 59}, 1 (1919).
\bibitem{sch}H. J. Schmidt, Int. J. Geom. Meth. Mod. Phys., {\bf 04}, 209-248 (2007).
\bibitem{str}A. A. Starobinsky, Phys. Lett. B, {\bf 91}, 99-102 (1980).

%DE and Unified cosmic expansion of f(R)
\bibitem{defr}S. Capozziello, S. Carloni, A. Troisi, arXiv: astro-ph/0303041;

D. N. Vollick, Phys. Rev. D, {\bf 68}, 063518 (2003);

S. Capozziello, V. F. Cardone, S. Carloni, A. Troisi, Int. J. Mod. Phys. D, {\bf 12}, 1969-1982 (2003);

S. M. Carroll, V. Duvvuri, M. Trodden, M. S. Turner, Phys. Rev. D, {\bf 70}, 043528 (2004).
\bibitem{unfr}S. Nojiri, S. D. Odintsov, Phys. Rev. D, {\bf 68}, 123512 (2003); Gen. Relt. Grav., {\bf 36}, 1765-1780 (2004); Phys. Lett. B, {\bf 657}, 238-245 (2007); Phys. Rept., {\bf 505}, 59-144 (2011).

%f(R) Reviews
\bibitem{dej}R. H. Dejrah, arXiv:2502.17519[gr-qc].
\bibitem{far}V. Faraoni, arXiv:0810.2602 [gr-qc].
\bibitem{soti}T. P. Sotiriou, V. Faraoni, Rev. Mod. Phys., {\bf 82}, 451 (2010).
\bibitem{noj1}S. Nojiri, S. D. Odintsov, Int. J. Geom. Meth. Mod. Phys., {\bf 04},115-146 (2007).
\bibitem{fel}A. De Felice, S. Tsujikawa, Living Rev. Relativity, {\bf 13}, 3 (2010).
\bibitem{noj2}S. Nojiri, S. D. Odintsov, Physics Reports, {\bf 505}, 59 (2011).

%wormhole
\bibitem{flm}L. Flamm, Phys. Z., {\bf 17}, 448 (1916).
\bibitem{er}A. Einstein, N. Rosen, Phys. Rev., {\bf 48}, 73 (1935).
\bibitem{vis}M. Visser, Lorentzian Wormholes: From Einstein to Hawking. Springer, Berlin (1997).
\bibitem{mt} M. S. Morris and K. S. Thorne, Am. J. Phys. {\bf 56}, 395 (1988).
\bibitem{mty}M. S. Morris, K. S. Thorne, U. Yurtsever, Phys. Rev. Lett., {\bf 61}, 1446 (1988).
\bibitem{hv}D. Hochberg, M. Visser, Phys. Rev. Lett., {\bf 81}, 746-749 (1998).


%eos motivation
\bibitem{abt}L. A. Anchordoqui, S. Perez Bergliaffa and D. F. Torres, Phys. Rev. D, {\bf 55} , 5226 (1997).
\bibitem{reza1}M. R. Mehdizadeh, M. K. Zangeneh and F. S. N. Lobo, Phys. Rev. D, {\bf 91}, 084004 (2015).
\bibitem{reza2}M. R. Mehdizadeh, F. S. N. Lobo, Phys. Rev. D, {\bf 93}, 124014 (2016).
\bibitem{kiro}S. Kiroriwal, J. Kumar, S. K. Maurya and S. Ray , Phys. Dark Univ., {\bf 46}, 101559 (2024).
\bibitem{zub}M. Zubair, M. Farooq, E. Gudekli, H. R. Kausar, and G. D. Acan Yildiz, Int. J. Geom. Methods Mod. Phys., {\bf 20}, 2350191 (2023).

\bibitem{buch}H. A. Buchdahl, Mon. Not. Roy. Astron. Soc., {\bf 150}, 1 (1970).


%f(R) gravity wh with nec
\bibitem{fr} N. Furey and A. De Benedictis, Class. Quant. Grav. {\bf 22}, 313, (2005);

F. S. N. Lobo, M. A. Oliveira, Phys. Rev. D, {\bf 81}, 067501 (2010);

F. S. N. Lobo, AIP Conf. Proc. {\bf 1458}, 447, (2011);

A. De Benedictis and D. Horvat, Gen. Rel. Grav. {\bf 44}, 2711, (2012);

T. Harko, F. S. N. Lobo, M. K. Mak and S. V. Sushkov, Phys. Rev. D {\bf 87}, 067504, (2013); 

P. Pavlovic, M. Sossich, Eur. Phys. J. C, {\bf 75} 117 (2015);

S. Bhattacharya, S. Chakrabarty, Eur. Phys. J. C {\bf 77 }, 558 (2017);

S. Bhattacharya, S. Halder and S. Chakraborty, Mod. Phys. Lett. A, {\bf 34}, 1950200 (2019).


%f(R) ghost
\bibitem{bron}K. A. Bronnikov, A. A. Starobinsky , JETP Lett., {\bf 85}, 1 (2007);

K. A. Bronnikov, M. V. Skvortsova and A. A. Starobinsky, Gravit. Cosmol., {\bf 16}, 216 (2010).

%solidangle-nonassymp
\bibitem{lob1}F. S. N. Lobo, Phys. Rev. D, {\bf 71}, 084011 (2005).
\bibitem{cat1}M. Cataldo, L. Liempi and P. Rodriguez, Phys. Lett. B, {\bf 757},130-135 (2016).
\bibitem{cat2}M. Cataldo, L. Liempi and P. Rodriguez, Eur. Phys. J. C, {\bf 77}, 748 (2017).
\bibitem{cat3}M. Cataldo and F. Orellana, Phys. Rev. D, {\bf 96}, 064022 (2017).
\bibitem{sg}R. Sengupta, S. Ghosh, M. Kalam and S. Ray, Classical Quant. Gravit., {\bf 39}, 105004 (2022).
\bibitem{rah2}F. Rahaman, B. Samanta, N. Sarkar, B. Raychaudhuri and B. Sen, Eur. Phys. J. C, {\bf 83}, 395 (2023).

%powerlaw f(r) for sol1 and 2
\bibitem{odi1}S. D. Odintsov and V. K. Oikonomou, Phys. Rev. D., {\bf 99}, 104070 (2019).
\bibitem{odi2}S. D. Odintsov and V. K. Oikonomou, Phys. Rev. D., {\bf 101}, 044009 (2020).
\bibitem{odi3}S. Nojiri, S. D. Odintsov and V. K. Oikonomou, Phys. Dark Univ., {\bf 29}, 100602 (2020).
\bibitem{odi4}S. D. Odintsov, D. Saez-Chillon Gomez and G. S. Sharov, Nucl. Phys. B, {\bf 966} 115377 (2021).
\bibitem{odi5}S. D. Odintsov, V.K. Oikonomou and G. S. Sharov, Phys. Lett. B, {\bf  843}, 137988 (2023).
\bibitem{sham1}M. F. Shamir, The Fifteenth Marcel Grossmann Meeting, pp. 459-464 (2022). $https://doi.org/10.1142/9789811258251_0057$
\bibitem{gog}D.J. Gogoi, U. D. Goswami, Indian J Phys., {\bf 96}, 637–646 (2022).
\bibitem{inf}V. Muller, H-J Schimdt and A. A. Staobinsky, Classical Quant. Gravit., {\bf 7}, 1163 (1990);

A. K. Sharma and M. M. Verma, ApJ, {\bf 926}, 29 (2022).

%sol4
\bibitem{cb}T. Clifton, J. D. Barrow, Phys. Rev. D, {\bf 72}, 103005 (2005).
\bibitem{clf}S. Copzziello, M. De Laurentis, M. Francaviglia, Astropart. Phys. Rev. D, {\bf 29}, 125-129 (2008).
\bibitem{cap1}S. Capozziello, A. Stabile, A. Troisi, Classical Quantum Gravit. {\bf 25}, 085004 (2008);

S. Capozziello, M. De Laurentis, A. Stabile, Classical Quantum Gravit. {\bf 27}, 165008 (2010).
\bibitem{mrf}M.De Laurentis, R. De Rosa, F. Garufi, L. Milano, MNRAS, {\bf 424}, 2371-2379 (2012).
\bibitem{cap2}S. Capozziello, M. De Laurentis, R. Farinelli, S. D. Odintsov, Phys. Rev. D, {\bf 93}, 023501 (2016);

A. V. Astashanok, S. Capozziello, S. D. Odintsov, V. K. Oikonomou, Phys. Lett. B, {\bf 811}, 135910 (2020).
\bibitem{cap3}S. Capozziello, O. Luongo, L. Mauro, Eur. Phys. J. Plus, {\bf 136}, 167 (2021).

\bibitem{cap5}E. Battista, S. Capozziello, A. Errehymy, Eur. Phys. J. C, {\bf 84}, 1314 (2024).

%nec in other gravity
\bibitem{gbll} G. Dotti, J. Oliva and R. Troncoso, Phys. Rev. D {\bf 75}, 024002 (2007);

G. Dotti, J. Oliva and R. Troncoso, Phys. Rev. D {\bf 76}, 064038 (2007);

H. Maeda and M. Nozawa, Phys. Rev. D {\bf 78}, 024005 (2008);

G. Dotti, J. Oliva and R. Troncoso, Int. J. Mod. Phys. A {\bf 24}, 1690 (2009);

A. B. Balakin, J. P. S. Lemos and A. E. Zayats, Phys. Rev. D \textbf{81}, 084015 (2010);

J. Matulich and R.Troncoso, J. High Energy Phys. {\bf 10}, 118 (2011).

\bibitem{rs} M. La Camera, {\it Phys. Lett. B}, {\bf 573}, 27, (2003).
\bibitem{rast}S. Halder, S. Bhattacharya and S. Chakraborty, Mod. Phys. Lett. A, {\bf 34}, 1950095 (2019).


%bd gravity
\bibitem{chb}T. Chiba, Phys. Lett. B, {\bf 575}, 1-3 (2003).
\bibitem{bd}C. Brans C R. H. Dicke, Phys. Rev., {\bf 124}, 925–35 (1961).

%Cauchy Problem
\bibitem{cp}W. J. Cocke and J. M. Cohen, J. Math. Phys., {\bf 9}, 971, (1968); 

D. R. Noakes, J. Math. Phys., {\bf 24}, 1846, (1983);

P. Teyssandier, Ph. Tourrenc, J. Math. Phys., {\bf 24}, 2793, (1983).

%bd-f(R) conversion
\bibitem{no}S. Nojiri, S. D. Odintsov, Phys. Rev. D, {\bf 68}, 123512 (2003).
\bibitem{st1}T. P. Sotiriou, Class. Quantum Grav., {\bf 23}, 5117 (2006).
\bibitem{st2}T. P. Sotiriou and V. Faraoni, Rev. Mod. Phys., {\bf 82}, 451(2010).

%bd wormhole
\bibitem{ag}A. G. Agnese and M. La Camera, Phys. Rev. D, {\bf 51}, 2011 (1995).
\bibitem{bdwm}K. K. Nandi, A. Islam, and J. Evans, Phys. Rev. D, {\bf 55}, 2497 (1997);

L. Anchordoqui, S. P. Bergliaffa, and D. F. Torres, Phys. Rev. D, {\bf 55}, 5226 (1997);

K. K. Nandi, B. Bhattacharjee, S. M. K. Alam, and J. Evans, Phys. Rev. D, {\bf 57}, 823 (1997);

A. Bhattacharya, I. Nigmatzyanov, R. Izmailov, and K. K. Nandi, Classical Quantum Gravit., {\bf 26}, 235017 (2009).
\bibitem{lb}F. S. N. Lobo and M. A. Oliveira, Phys. Rev. D, {\bf 81}, 067501 (2010).
\bibitem{cap4}S. Capozziello, S. Nojiri, S. D. Odintsov, Phys. Lett. B, {\bf 634}, 93-100, (2006).
\bibitem{vf}V. Faraoni and S. D. Belknap-Keet, Phys. Rev. D, {\bf 96}, 044040 (2017).

%Hybrid-Palatini
\bibitem{harko}T. Harko, T. S. Koivisto, F. S. N. Lobo and G. J. Olmo, Phys. Rev. D, {\bf 85}, 084016 (2012);

S. Capozziello, T. Harko, F. S. N. Lobo and G. J. Olmo, Int. J. Mod. Phys. D, {\bf 22}, 1342006 (2013).

\bibitem{cap6}S. Capozziello, T. Harko, T. S. Koivisto, F. S. N. Lobo, and G. J. Olmo, Phys. Rev. D, {\bf 86}, 127504 (2012).

%cosmological viability
\bibitem{stfr}A. D. Dolgov, M. Kawasaki, Phys. Lett. B, {\bf 573}, 1-4 (2003);

G. J. Olmo, Phys. Rev. D, {\bf 72}, 083505 (2005).
\bibitem{chiba}T. Chiba, Phys. Lett. B, {\bf 575}, 1-3 (2003).
\bibitem{vifr}V. Faraoni, Phys. Rev. D, {\bf 74}, 023529 (2006);

I. Navarro, K. Van Acoleyn, JCAP, {\bf 0702}, 022 (2007);

L. Amendola, R. Gannauji, D. Polarski, S. Tsujikawa, Phys. Rev. D, {\bf 75}, 083504 (2007);

Yong-Seon Song, W. Hu, I. Sawicki, Phys. Rev. D, {\bf 75}, 044004 (2007);

L. Pogosian, A. Silvestri, Phys. Rev. D, {\bf 77}, 023503 (2008).


%Complexity and lr
\bibitem{raj}R. Shaikh, P. Banerjee, S. Paul, T. Sarkar, Phys. Lett. B, {\bf 789}, 270-275 (2019); {\it Erratum}, 791, 422-423 (2019).
\bibitem{her1}L. Herrera Phys. Rev. D, {\bf 97}, 044010 (2018);

L. Herrera, A. Di Prisco and J. Ospino, Phys. Rev. D, 98, 104059 (2018).
\bibitem{her2}L. Herrera, A. Di Prisco, J. Ospino, Gen. Relativ. Gravity, {\bf 42}, 1585–1599 (2010).
\bibitem{snsb1}S. Bhattacharya, S. Nalui, J. Math. Phys., {\bf 64}, 052501 (2023) arXiv:2304.08877 [gr-qc].
\bibitem{snsb2}S. Nalui, S. Bhattacharya, Ann. Phys., {\bf 470}, 169789 (2024).
\bibitem{snsb3}S. Nalui, S. Bhattacharya, Phys. Lett. B, {\bf 861}, 139261 (2025).




\end{thebibliography}
\end{document}